\documentclass[structabstract]{aa}  
\usepackage{graphicx}
\usepackage{txfonts}
\usepackage{longtable}
\usepackage{lscape}
\usepackage{natbib}
\usepackage{color}
\bibpunct{(}{)}{;}{a}{}{,} % to follow the A&A style
\begin{document}
   \title{The Penn State--Toru\'n Centre for Astronomy Planet Search stars
	\thanks{Based on observations obtained with the Hobby-Eberly Telescope, which is a joint project of the University of Texas at Austin, the 		Pennsylvania State University, Stanford University, Ludwig-Maximilians-Universit\"at M\"unchen, and Georg-August-Universit\"at G\"ottingen.}
	}

   \subtitle{I. Spectroscopic analysis of 348 red giants}

   \author{P. Zieli\'nski
          \inst{1}
          \and
	   A. Niedzielski
          \inst{1}
          \and
           A. Wolszczan
	  \inst{2,3}
          \and
	   M. Adam\'ow
          \inst{1}
          \and
	   G. Nowak
          \inst{1}
          }

   \institute{Toru\'n Centre for Astronomy, Nicolaus Copernicus University, Gagarina 11, 87-100 Toru\'n, Poland\\
              \email{pawziel@astri.umk.pl, aniedzi@astri.umk.pl, adamow@astri.umk.pl, grzenow@astri.umk.pl}
         \and
             Department of Astronomy and Astrophysics, Pennsylvania State University, 525 Davey Laboratory, University Park, PA 16802
              \email{alex@astro.psu.edu}
	 \and
             Center for Exoplanets and Habitable Worlds, Pennsylvania State University, 525 Davey Laboratory, University Park, PA 16802
             }

   \date{Received ; accepted }

% \abstract{}{}{}{}{} 
% 5 {} token are mandatory
 
  \abstract
  % context heading (optional), leave it empty if necessary  
{}
{We present basic atmospheric parameters ($T_{\rm{eff}}$, $\log g$, $v_{\rm{t}}$ and [Fe/H]) as well as luminosities, masses, radii and absolute radial velocities for 348 stars, presumably giants, from the $\sim$ 1000 star sample observed within the Penn State-Toru\'n Centre for Astronomy Planet Search with the High Resolution Spectrograph of the 9.2~m Hobby-Eberly Telescope. The stellar parameters (luminosities, masses, radii) are key ingredients in proper interpretation of newly discovered low-mass companions while a systematic study of the complete sample will create a basis for future statistical considerations concerning low-mass companions appearance around evolved low and intermediate-mass stars.
}
{The atmospheric parameters were derived using a strictly spectroscopic method based on the LTE analysis of equivalent widths of Fe\,I and Fe\,II lines. With existing photometric data and the Hipparcos parallaxes we estimated stellar masses and ages via evolutionary tracks fitting. The stellar radii were calculated from either estimated masses and the spectroscopic $\log g$ or from the spectroscopic $T_{\rm{eff}}$ and estimated luminosities. The absolute radial velocities were obtained by cross-correlating spectra with a numerical template. 
}
{We completed the spectroscopic analysis for 332 stars of which 327 were found to be giants. For the remaining 16 stars with incomplete data a simplified analysis was applied. The results show that our sample is composed of stars with effective temperatures ranging from 4055 K to 6239 K, and $\log g$ between 1.39 and 4.78 (5 dwarfs were identified). The estimated luminosities ranging between $\log L/L_{\odot} = -1.0$ and 3 lead to masses ranging from 0.6 to 3.4~$M_{\odot}$. Only 63 stars with masses larger than 2~$M_{\odot}$ were found. The radii of our stars range from 0.6 to 52~$R_{\odot}$ with vast majority between 9-11~$R_{\odot}$. The stars in our sample are generally less metal abundant than the Sun with median [Fe/H] $=-0.15$. The estimated uncertainties in the atmospheric parameters were found to be comparable to those reached in other studies. However, due to lack of precise parallaxes the stellar luminosities and, in turn, the masses are far less precise, within 0.2~$M_{\odot}$ in best cases, and 0.3~$M_{\odot}$ on average.
}
{}


   \maketitle
%
%________________________________________________________________

\section{Introduction}

Since the discovery of the first extrasolar planetary systems by \citet{WF1992,MQ1995} and \citet{MB1996} over 700 planets were found around other stars. The richness of exoplanets and the architectures of planetary systems that are today emerging is amazing and rises questions about a general picture of planet formation and evolution. Before such a general picture will be available, studies of planetary systems in various environments are essential. 

The observational techniques applied in searches for exoplanets are all limited to their areas of competence -- a range in stellar or planetary parameters (effective temperatures, masses, planetary periods or orbital separations etc.) that are available to them. Due to the current detectors limitations the direct imaging searches for exoplanets deliver typically massive planets at large orbital distances from low-mass stars (cf. 2M1207 b -- \citealt{Chau2004} or HD~8799 system -- \citealt{Mar2008}). The transit planet searches, conveniently operating in the $\sin i \approx$ 1 domain ($i$ being orbital inclination), discover exoplanets with relatively short orbital periods (orbital separations) due to the natural limit of the observable transit length set by the lengths of a night (a condition now being relaxed with the multi-site or space observations, cf. HD~80606 b -- \citealt{Hid2010}, \citealt{Heb2010}, Kepler~11g -- \citealt{Lis2011}). The searches for planets based on the radial velocity (RV) technique are the most versatile. They detect planetary candidates at various orbital separations and in a wide range of planetary minimum masses, as long as $\sin i \ne \rm{0}$ (cf. for example GJ~581 system -- \citealt{May2009} and 47~UMa d -- \citealt{GF2010}). Even this method has its limitations, though. Since it requires a large number of narrow spectral lines it focuses on the slowly rotating cool stars of late-F to M spectral type what sets an upper limit to the Main Sequence planetary hosts masses ($M \lesssim 1.5 M_{\odot}$). This technique is unfortunately also seriously affected by the stellar activity \citep{Que2001} which, if not resolved, adds noise (jitter) and limits the precision in RV \citep{Car2003, Hek2006, Cumm2008}. It is, however the most suitable in searches for planets around stars more massive than $\sim1.5 M_{\odot}$. In turn of evolution these stars slow down their rotation and decrease their effective temperatures becoming red giants available for precise RV measurements. 

The Penn State-Toru\'n Centre for Astronomy Planet Search (PTPS) is aiming at detection and characterization of planetary systems around stars at various evolutionary stages \citep{NW2008} with the long-term goal of description of the evolution of planetary systems with aging stars. To this end over 1000 stars are monitored with the Hobby-Eberly Telescope (HET) for RV variations using the high precision iodine-cell technique \citep{But1996}. The sample is mainly composed of evolved low-mass and intermediate-mass stars: $\sim575$ giants (including presented here 348 stars from the Red Giant Clump (RGC) sample) and $\sim225$ sub-giants but it also contains $\sim200$ slightly evolved dwarfs. Several stars with planetary-mass companions were already discovered, mainly around red giants \citep{Nie2007,Nie2009a,Nie2009b,Get2012a,Get2012b}. These findings supplemented the slowly growing population of $\sim50$ planets currently known around evolved stars, discovered in such projects like: the McDonald Observatory Planet Search \citep{CH1993,HC1993}, the Okayama Planet Search \citep{Sato2003}, the Tautenburg Planet Search \citep{Hat2005}, the Lick K-giant Survey \citep{Frink2002}, the ESO FEROS planet search \citep{Set2003b,Set2003a}, the Retired A Stars and Their Companions \citep{Joh2007}, the CORALIE $\&$ HARPS search \citep{LM2007}, the Boyunsen Planet Search \citep{Lee2011} and several others.

With this paper we start a series devoted to a detailed description of stars incorporated in the complete PTPS sample, their atmospheric parameters, elemental abundances, rotation velocities, kinematical properties, masses, radii, ages etc. The derived parameters will not only describe our sample but also will allow for proper interpretation of the planet search results. The stellar masses are essential in extracting the planetary mass companions minimum masses ($m_{\rm{P}} \sin i$), the radii and projected rotation velocities are key ingredients in the stellar activity considerations etc. Since the complete sample contains $\sim1000$ stars randomly distributed in the skies it is of interest for general stellar and galactic astrophysics as well. It may be used for studies of galactic structure and evolution, galactic distribution of planetary systems \citep{Hay2009} and the galactic habitable zone \citep{Gon2001, Line2004, McCL2010, Gow2011}. 

The purposes of present paper is to determine atmospheric parameters, such as effective temperatures ($T_{\rm{eff}}$), stellar gravitational accelerations ($\log g$), microturbulence velocities ($v_{\rm{t}}$) and metallicities ([Fe/H]) for 348 GK-type stars, presumably Red Clump giants observed within the PTPS survey. Together with existing photometric data and parallaxes (when available) they will allow us to estimate stellar masses ($M/M_{\odot}$), radii ($R/R_{\odot}$) and ages. 

The RGC, together with the Main Sequence (MS) and the red giant branch (RGB) is one of the most characteristic features of the Hertzsprung-Russell diagram (HRD). It was identified by \citet{Can1970} and \citet{Fau1973} as the location of the steady core helium burning. That finding was elaborated in detail by \citet{Gir1999}. \citet{Jim1998} used the RGC stars to estimate the age of the Galaxy. Due to very similar intrinsic brightness of all stars at that stage the RGC was proposed as a standard candle in distance estimates and applied to the Galactic Center \citep{Pacz1998}, the Large Magellanic Cloud \citep{Salar2003} and other stellar systems. 

Interesting stellar evolution phase and general astrophysical application make the red giants, and especially the RGC stars subject of numerous surveys. \citet{McW1990} obtained $T_{\rm{eff}}$ for 671 stars from broad-band photometry, $\log g$ via isochrone fitting as well as the local thermodynamical equilibrium (LTE) abundances from high resolution spectra. \citet{Zh2001} delivered atmospheric parameters, iron abundances, $\alpha$-element enhancements, and masses for 39 Red Clump giants selected from the Hipparcos catalogue. \citet{Fam2005} provided the CORAVEL radial velocities for a sample of about 6\,600 K-type giants. Detailed spectroscopic studies of various subsamples of the galactic Red Clump giants were presented in \citet{Mish2006}, \citet{Biz2006,Biz2010}, \citet{Hek2007}, \citet{Liu2007}, \citet{LH2007}, \citet{Tak2008}, \citet{Puz2010}. 

\citet{Gon2008} summarized kinematical properties and galactic distribution of 97\,348 RGC stars based on Tycho-2 proper motions and Tycho-2 and 2MASS photometry. \citet{Val2010} studied 277 RGC stars and presented accurate multi-epoch radial velocities, atmospheric parameters, distances and space velocities ($U, V, W$). Moreover, \citet{Tau2010} in a recent paper presented $^{12}$C/$^{13}$C abundance ratio determinations in a sample of 34 Galactic clump giants as well as abundances of nitrogen, carbon and oxygen. 

Meanwhile a new insight of giants, including clump giants, became available through CoRoT \citep{Bag2006} and Kepler \citep{Gil2010} precise photometry. The solar-like oscillations in red giants were first identified in ground-based RV observations \citep{Hat1994,Mer1999}. \citet{Rid2009} reported the presence of radial and non-radial oscillations in over 300 giants observed with CoRoT. These oscillations were used in massive masses and radii determinations for giants \citep{Kal2010,Bed2010,Hek2011}.

This paper is organized as follows. In Sect.~2, we describe the sample and the observational material. The procedure of Fe lines equivalent widths measurements and its precision are presented in Sect.~3. The outline of the method applied in our spectroscopic analysis is given in Sect.~4 while in the following Sect.~5 we present several tests to evaluate reliability of the method and estimate uncertainties. Section~6 contains summary of the determined atmospheric parameters for our stars. These parameters are used in Sect.~7 to estimate stellar integrated parameters: luminosities, masses, radii, ages. Section~8 presents the method and results of our RV determinations. The comparison of our results with other published data, whenever available, is presented in Sect.~9. Section~10 contains discussion of atmospheric parameters and conclusions are presented in Sect.~11.

\section{Targets selection and observations}

The sample presented here, 348 stars, originally the SIM-EPIcS\footnote{Space Interferometry Mission (SIM) key project: Extra-solar Planet Interferometric Survey (EPIcS)\\ http://planetquest.jpl.nasa.gov/SIM} reference candidates, selected as presumably RGC stars basing on their existing photometry and reduced proper motions \citep{Gel2005,Law2006} constitute the so-called RGC subsample of the PTPS. 

\begin{figure}
   \centering
   \includegraphics[angle=0,width=9cm]{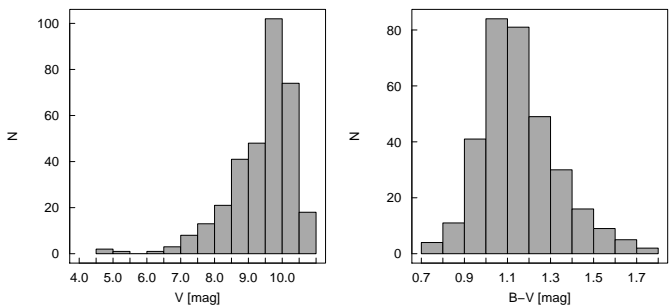}
      \caption{Histograms of the apparent magnitudes in V band (left panel) and (B$-$V) (right panel) for the 348 Red Clump giants observed within the PTPS.}
         \label{fig-obs}
   \end{figure}

They are relatively bright field stars, with V between 4.5 and 11~${\rm{mag}}$ but in the SIMBAD\footnote{The SIMBAD astronomical database is operated at CDS, Strasbourg, France.} database spectral types and/or luminosity class estimates are available only for a few (4 are classified as dwarfs and 22 as giants). The basic photometric data for the sample stars come from the 2MASS Point Source catalogue \citep{Cut2003} and the Tycho-2 catalogue \citep{Hog2000} compiled by \citet{Gel2005}. For several stars missing data were gathered from \citet{Khar2009}. The distrubitions of visual magnitudes and B$-$V color indices are presented in Fig.~\ref{fig-obs}. Vast majority, 92\% of stars, are fainter than V $=8~{\rm{mag}}$ with (B$-$V) between 0.7 to 1.8 magnitudes (1.1~${\rm{mag}}$ on average). 

Our sample is composed of much fainter stars than included in most of the other RGC surveys \citep{Fam2005,Mish2006,Hek2007,Tak2008,Val2010} and can be compared to the more numerous sample of \citet{Biz2006,Biz2010}. The list of stars, their identification and basic observational parameters are presented in Table~\ref{tab-cat}.

Our high quality, high resolution optical spectra were collected since 2004 with the HET \citep{Ram1998} located in the McDonald Observatory. The telescope was equipped with the High Resolution Spectrograph (HRS) fed with a 2~arcsec fiber, working in the R = 60\,000 resolution \citep{Tull1998}. The observations were performed in the queue scheduled mode \citep{Shet2007}. The spectra consisted of 46 echelle orders recorded on the ,,blue'' CCD chip ($407-592$~nm) and 24 orders on the ,,red'' one ($602-784$~nm). The signal to noise ratio was typically better than $200-250$ per resolution element. The basic data reduction (flat fielding, extraction of the spectrum, wavelength calibration, normalization to continuum) were performed using standard IRAF\footnote{IRAF is distributed by the National Optical Astronomy Observatories, which are operated by the Association of Universities for Research in Astronomy, Inc., under cooperative agreement with the National Science Foundation.} tasks and scripts. The wavelength calibration was based on Th-Ar lamp spectrum identification presented in \citet{MS2002}. 

\addtocounter{table}{1}

\section{Equivalent widths of Fe lines}

   \begin{figure}
   \centering
   \includegraphics[angle=-90,width=8cm]{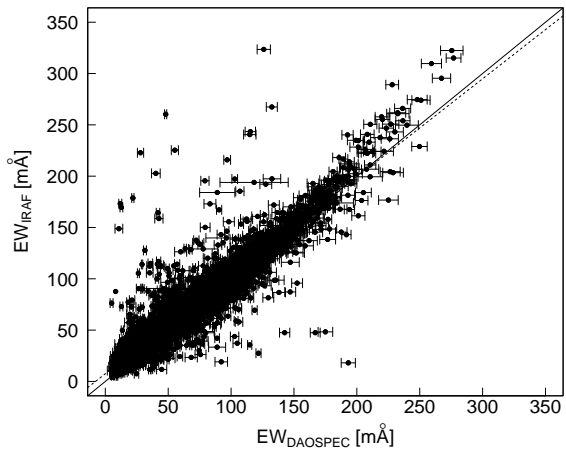}
      \caption{A comparison of the automatic (DAOSPEC) and hand-made (IRAF) EWs measurements of 5\,695 Fe lines for 29 stars. The solid line presents the one to one relation, the dashed one is the best linear fit to the data. The individual uncertainties of the DAOSPEC EWs measurements are presented.}
         \label{fig-ew}
   \end{figure}

Since equivalent widths (EWs) of hundreds of iron lines for every object were needed we used the DAOSPEC\footnote{http://www2.cadc-ccda.hia-iha.nrc-cnrc.gc.ca/community/STETSON/daospec\\ http://www.bo.astro.it/$\sim$pancino/projects/daospec.html} \citep{SP2008} to measure the EWs in an automatic manner. However, to validate reliability of the results and estimate uncertainties in EWs we measured spectra of 29 stars manually with the IRAF (Gaussian fitting) as well. In Fig.~\ref{fig-ew} the comparison of the IRAF (manual) and the DAOSPEC (automatic) measurements is presented. We found that the best agreement between the measurements is reached for lines with the EWs lower than 200~m\AA. Lines with larger EWs revealed some discrepancy towards higher manual EWs values. In general, the linear fit to all measurements is nearly a one to one relation. For 5\,632 lines with the EW $\leq$ 200~m\AA ~we found EW$_{\rm{IRAF}}=(0.937\pm0.005)$EW$_{\rm{DAOSPEC}}+(8.46\pm0.35)$ and the Pearson correlation coefficient of $R=0.939$. The mean difference between the IRAF and the DAOSPEC measurements was 5.3~m$\AA$ ~with a scatter of 12.8~m\AA. ~The mean uncertainty in the DAOSPEC EWs was 2.13~m\AA.~We proceeded with the automatic measurements in the further analysis. 

To apply the procedure described in the following section we measured the EWs of neutral (Fe\,I) and ionized (Fe\,II) iron absorption lines for which precise atomic data, i.e. laboratory wavelenghts, lower excitation potentials ($\chi$) and oscillator strength ($g f$) were available from \citet{Tak2005a,Tak2005b}. All these atomic data were compiled from the line-lists of \citet{GS1999,Mey1993} and \citet{Kur1984}. Due to differences in the spectral coverage and the CCD chip cosmetics we were able to identify up to 296 lines (269 Fe\,I and 27 Fe\,II lines) from the initial sample of $\sim330$ iron lines from \citet{Tak2005a} in the HET/HRS spectra. After rejecting blends, very weak and saturated lines we used the EWs of 190 Fe\,I and 15 Fe\,II lines per star on average. Only unblended lines with the EWs ranging from 5~m\AA ~to 200~m\AA ~were included in the final spectroscopic analysis.

\section{An outline of the analysis method}

The iron EWs were analyzed with the TGVIT \citep{Tak2002a,Tak2005a}, which is a part of the SPTOOL{\footnote{http://optik2.mtk.nao.ac.jp/$\sim$takeda/sptool}} package. This numerical algorithm is a modified version of an earlier code and is based on the atmospheric models computed by \citet{Kur1993a,Kur1993b}. The TGVIT uses an iterative procedure to achieve more cohesive parameter fits and in the version available to us was suitable for lower effective temperatures ($T_{\rm{eff}}$ $\geq$ $4250$~K) than the original version \citep{Tak2002a}, so we could apply it to the late-type stars in our sample. More details concerning the iteration procedures, data calculations and formulation of the numerical problems are presented in \citet{Tak2002a, Tak2002b, Tak2005a, Tak2005b}.

This purely spectroscopic method is based on analysis of Fe\,I and Fe\,II lines and relies on three conditions resulting from the assumption of the local thermodynamical equilibrium (LTE) that have to be satisfied:
\begin{enumerate}
\item The abundances derived from Fe\,I lines may not show any dependence on the lower excitation potential $\chi$ (excitation equilibrium);
\item The averaged abundances from Fe\,I and Fe\,II lines must be equal (ionization equilibrium);
\item The abundances derived from Fe\,I lines may not show any dependence on the equivalent widths EWs (matching the curve of growth shape).
\end{enumerate}

We used the TGVIT to calculate $T_{\rm{eff}}$, $\log g$, $v_{\rm{t}}$ and [Fe/H], defined in the standard manner as [Fe/H]=$\log(N_{\rm{Fe}}/N_{\rm{H}})-\log(N_{\rm{Fe}}/N_{\rm{H}})_{\odot}$, with the solar value of 7.50 \citet{Kur1993a,Hol1991}. 

   \begin{figure}
   \centering
   \includegraphics[angle=-90,width=9.5cm]{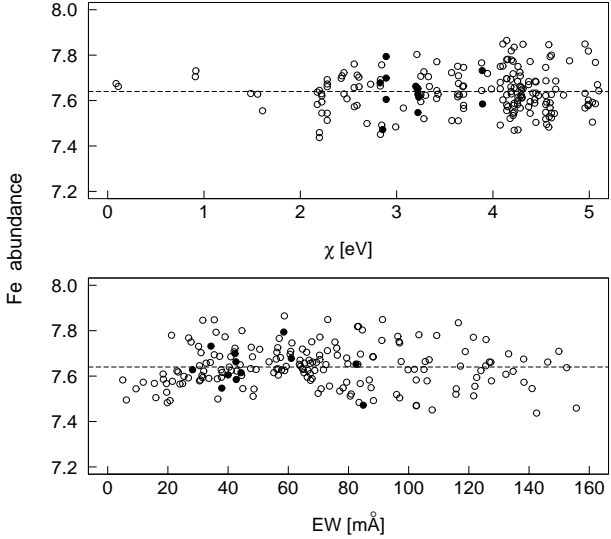}
      \caption{An illustration of the TGVIT analysis of TYC 3018-00996-1: Fe abundance vs. $\chi$ (top panel) relation and Fe abundance vs. EW (bottom panel) relation. The open and filled circles stand for 176 Fe\,I and 12 Fe\,II lines respectively. The mean Fe abundance is showed on each panel as a dashed line.}
         \label{fig-chiew}
   \end{figure}

Because of relatively low $T_{\rm{eff}}$ of our stars the number of available Fe\,II lines was not sufficiently large (less than 10\% of Fe\,I lines) and consequently we neglected the dispersion of Fe\,II abundances while estimating the minimization function $D^2$ (TGVIT defines the dispersion $D^2$ as a function of three arguments: $T_{\rm{eff}}$, $\log g$, $v_{\rm{t}}$ that is minimized to obtain proper Fe abundance). For the coefficients included in $D^2$ we assumed the values of: $c_1=0$, $c_2=1$, $c_3=0$ (we refer the reader to \citet{Tak2002a} for details). 

In the vast majority of cases all three above-mentioned conditions were fulfilled what was checked for every star by visual evaluation of relations between Fe abundances and $\chi$ or EWs like those presented in Fig.~\ref{fig-chiew}. The assumptions were justified if no relation between Fe abundance and these two parameters was present. Another conditions evaluated for every star were the equivalence of mean Fe abundances for Fe\,I and Fe\,II lines and the scatter around average abundance. In turn of visual inspection the Fe lines showing apparent deviations from the mean Fe abundances were rejected. In the following step the parameter were redetermined and the procedure was repeated, if necessary, until coherent solutions for all parameters were reached. The rejected lines were typically unrecognized blends or lines with EWs larger than 150~m\AA ~with the excitation potential of less than 0.5~eV.

The Fe abundance depends on both $T_{\rm{eff}}$ and $\log g$. The coupling is so strong that in the iterative approach of \citet{Tak2002a} $T_{\rm{eff}}$ and $\log g$ are actually simultaneously fitted and treated as ,,one variable'' ($v_{\rm{t}}$ being the other). This coupling may, in principle, lead to degeneracy of purely spectroscopic parameters as higher effective temperature may be compensated by the gravitational acceleration. However, in our sample, composed of stars with effective temperatures similar to the Sun or lower we do not expect this effect to introduce significant uncertainties. We refer the reader to \citet{Tak2002a} for a detailed discussion of this problem as well as for an extensive discussion on the behavior of the $D^2$ function, along with its related components, around the converged solutions.

\section{Verification of the adopted approach}

\begin{table*}
\caption{Atmospheric parameters of the Sun and $\tau$~Cet determined by \citet{Tak2005a} and our results.}
\label{tab-takeda} 
\centering 
\begin{tabular}{ll|rrrr|rrrr}
\hline\hline 
& & \multicolumn{4}{c|}{Takeda et al.} & \multicolumn{4}{c}{our results} \\
Name & Sp. & $T_{\rm{eff}}$ & $\log g$ & $v_{\rm{t}}$ & [Fe/H] & $T_{\rm{eff}}$ & $\log g$ & $v_{\rm{t}}$ & [Fe/H] \\
& Type & [K] & [cm s$^{-2}$]  & [km s$^{-1}$] & & [K] & [cm s$^{-2}$] & [km s$^{-1}$] & \\
\hline 
Sun	 		& G2V & 5785 $\pm$ 8 & 4.44 $\pm$ 0.02 & 0.96 $\pm$ 0.06 & 0.01 $\pm$ 0.04 & 5785 $\pm$ 8 & 4.44 $\pm$ 0.02 & 0.96 $\pm$ 0.06 & 0.02 $\pm$ 0.04 \\
$\tau$~Cet (HD 10700)	& G8V & 5420 $\pm$ 0 & 4.68 $\pm$ 0.00 & 0.66 $\pm$ 0.10 &$-$0.43 $\pm$ 0.06 & 5420 $\pm$ 0 & 4.68 $\pm$ 0.00 & 0.65 $\pm$ 0.10 &$-$0.43 $\pm$ 0.06 \\
\hline 
\end{tabular}
\end{table*}
\begin{table*}
\caption{Atmospheric parameters determined by \citet{Sou1998} and our results for 13 stars observed with ELODIE.}
\label{tab-elodie} 
\centering 
\begin{tabular}{ll|rrr|rrr}
\hline\hline 
& & \multicolumn{3}{c|}{Soubiran et al.} & \multicolumn{3}{c}{our results} \\
Name & Sp. & $T_{\rm{eff}}$ & $\log g$ &  [Fe/H] & $T_{\rm{eff}}$ & $\log g$ & [Fe/H] \\
& Type & [K] & [cm s$^{-2}$] &  & [K] & [cm s$^{-2}$] &  \\
\hline 
HD 10380	& K3III	& 4057 $\pm$ 74 & 1.43 $\pm$ 0.25 &$-$0.25 $\pm$ 0.14 & 4192 $\pm$ 13 & 2.03 $\pm$ 0.06 & $-$0.36 $\pm$ 0.07 \\
HD 10700	& G8V   & 5264 $\pm$ 74 & 4.36 $\pm$ 0.25 &$-$0.50 $\pm$ 0.14 & 5311 $\pm$ 28 & 4.40 $\pm$ 0.07 & $-$0.41 $\pm$ 0.10 \\
HD 22049  	& K2V   & 5052 $\pm$ 74 & 4.57 $\pm$ 0.25 &$-$0.15 $\pm$ 0.14 & 5113 $\pm$ 40 & 4.65 $\pm$ 0.10 &  0.04 $\pm$ 0.12 \\
HD 30834	& K3III & 4115 $\pm$ 74 & 1.73 $\pm$ 0.25 &$-$0.21 $\pm$ 0.14 & 4246 $\pm$ 15 & 1.99 $\pm$ 0.08 & $-$0.45 $\pm$ 0.08 \\	
HD 34411	& G0V   & 5835 $\pm$ 63 & 4.17 $\pm$ 0.20 & 0.06 $\pm$ 0.14 & 5886 $\pm$ 15 & 4.20 $\pm$ 0.03 &  0.15 $\pm$ 0.05 \\
HD 39003	& K0III & 4610 $\pm$ 74 & 2.18 $\pm$ 0.25 &$-$0.10 $\pm$ 0.14 & 4618 $\pm$ 30 & 2.18 $\pm$ 0.11 & $-$0.03 $\pm$ 0.14 \\
HD 39853	& K3	& 3994 $\pm$ 74 & 1.00 $\pm$ 0.25 &$-$0.40 $\pm$ 0.14 & 4094 $\pm$ 13 & 1.63 $\pm$ 0.06 & $-$1.00 $\pm$ 0.08 \\
HD 113226	& G8III & 4983 $\pm$ 74 & 2.80 $\pm$ 0.25 & 0.05 $\pm$ 0.14 & 5016 $\pm$ 28 & 2.49 $\pm$ 0.09 &  0.11 $\pm$ 0.13 \\
HD 124897	& K2III & 4361 $\pm$ 74 & 1.93 $\pm$ 0.25 &$-$0.53 $\pm$ 0.14 & 4305 $\pm$ 15 & 1.77 $\pm$ 0.07 & $-$0.55 $\pm$ 0.09 \\
HD 132142	& K1V	& 5108 $\pm$ 74 & 4.50 $\pm$ 0.25 &$-$0.55 $\pm$ 0.14 & 5170 $\pm$ 15 & 4.64 $\pm$ 0.04 & $-$0.35 $\pm$ 0.07 \\
HD 175305	& G5III & 4899 $\pm$ 131 & 2.30 $\pm$ 0.43 &$-$1.43 $\pm$ 0.20 & 5074 $\pm$ 30 & 2.89 $\pm$ 0.16 & $-$1.31 $\pm$ 0.12 \\
HD 204613	& G0III & 5727 $\pm$ 63 & 3.84 $\pm$ 0.20 &$-$0.51 $\pm$ 0.14 & 5842 $\pm$ 18 & 4.13 $\pm$ 0.04 & $-$0.21 $\pm$ 0.06 \\
HD 217014	& G5V	& 5729 $\pm$ 63 & 4.12 $\pm$ 0.20 & 0.11 $\pm$ 0.14 & 5789 $\pm$ 18 & 4.34 $\pm$ 0.05 &  0.24 $\pm$ 0.07 \\
\hline 
\end{tabular}
\end{table*}

Before running the TGVIT for our program stars we performed a series of test using various input data. First of all we tested our installation of the TGVIT by running it with the data for the Sun and $\tau$~Cet (HD~10700) distributed together with the code. These EWs were obtained from high dispersion spectra obtained at the Okayama Astrophysical Observatory with the 188~cm telescope \citep{Tak2005a}. Our results are compared with those of \citet{Tak2005a} in Table~\ref{tab-takeda}. The results are practically identical which proves that our installation works well. 

In a following step we determined atmospheric parameters (i) for several stars, for which similar determinations were available form \citet{Katz1998} using Echelle spectra available from the ELODIE Library \citep{Sou1998} and (ii) for a few well-known stars with planets, for which recent atmospheric parameters determinations are available in \citet{But2006}, using our own HET/HRS spectra. The purpose of these tests was to estimate reliability of the adopted method and to search for systematic effects. 

Another important reason for that analysis was to assess reliability of uncertainties delivered by the TGVIT as these are determined from the quality of fits for ${T_{\rm{eff}}}$, $\log g$ and microturbulence velocity and should be considered as intrinsic. Only the uncertainties in [Fe/H] are based on actual scatter between individual determinations for all lines in use and are expected to be realistic.

   \begin{figure*}
   \centering
   \includegraphics[angle=-90,width=16cm]{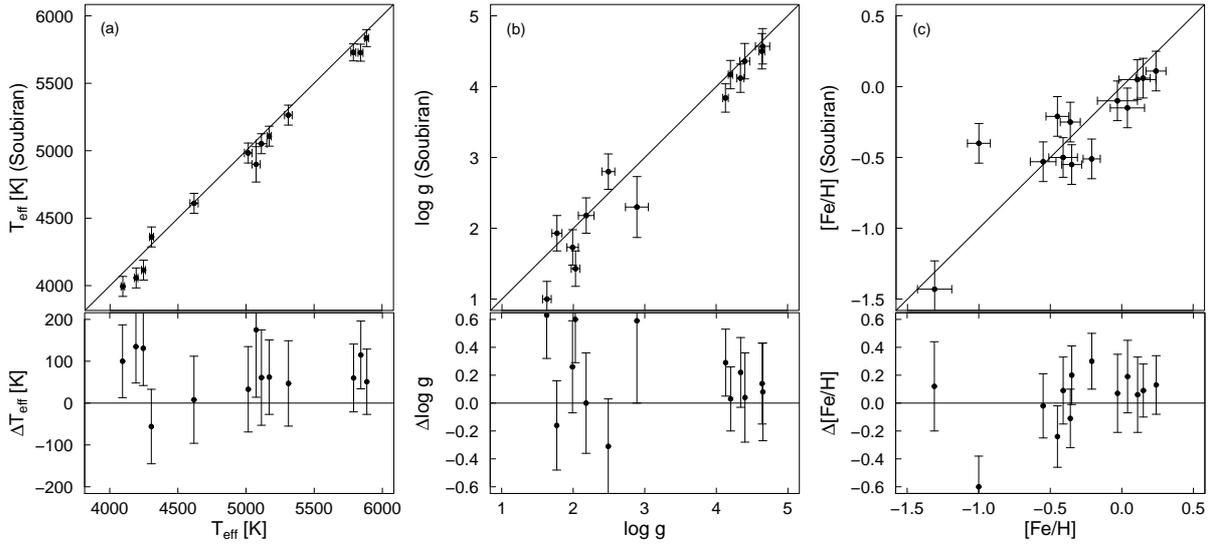}\vspace{-2cm}
      \caption{A comparison of the atmospheric parameters for 13 stars observed with the ELODIE. In the following panels the relations (top) and differences (bottom) between our results and those obtained by \citet{Sou1998} for $T_{\rm{eff}}$ (a), $\log g$ (b) and [Fe/H] (c) are presented. The solid lines represent one to one relations.}
         \label{fig-elodie}
   \end{figure*}

\subsection{Tests with ELODIE data}

We analyzed with the TGVIT spectra of randomly selected 13 stars downloaded from the ELODIE Library \citep{Sou1998}\footnote{Library of the ELODIE spectra of F5 to K7 stars is available only in electronic form via \\ http://vizier.cfa.harvard.edu/viz-bin/VizieR?-source=J/A+AS/133/221} at the Observatoire de Haute-Provence. Both giants and dwarfs with spectral types of G0-K3 were selected to cover similar parameter space to our sample.

These spectra were obtained with lower than ours resolution of R = 42\,000 \citep{Katz1998}. The spectral wavelength range covered was $390-680$~nm in 67 Echelle orders. Typical signal to noise ratio was $100-150$ per pixel. The spectra were reduced, i.e. straightened, wavelength calibrated, cleaned from cosmic ray hits, bad pixels and telluric lines. Because of the narrower spectral range we were able to measure 150 Fe\,I and 15 Fe\,II lines per star on average for the further analysis. 

The results obtained with the TGVIT for these 13 stars are presented and compared with results of \citet{Sou1998} in Table~\ref{tab-elodie} and 
Fig.~\ref{fig-elodie}. 

The mean intrinsic uncertainties of our determinations are $\sigma{T_{\rm{eff}}}=21$~K, $\sigma{\log~g}=0.07$ and $\sigma$[Fe/H] $=0.09$, respectively. According to \citet{Katz1998}, the average uncertainties based on the internal accuracy assumed for the stars with signal to noise ratio around 100 are $\sigma{T_{\rm{eff}}}=86$~K, $\sigma{\log~g}=0.28$ and $\sigma$[Fe/H] $=0.16$. We note that the mean intrinsic uncertainties in ${T_{\rm{eff}}}$ and $\log~g$ as given by the TGVIT are about 4 times smaller than uncertainties estimated by \citet{Katz1998} while in the case of [Fe/H] they are comparable. An overall agreement of the results is clear from Fig.~\ref{fig-elodie}.

We calculated the mean differences and scatter (rms) between our results and those of \citet{Sou1998} (excluding HD 175305 for which uncertainties in \citet{Sou1998} are much larger than typical) and obtained: $\Delta{T_{\rm{eff}}}=63$~K, $\Delta{\log~g}=0.15$, $\Delta$[Fe/H] $=0.02$, $\sigma_{\rm{D}}{T_{\rm{eff}}}=54$~K, $\sigma_{\rm{D}}{\log~g}=0.27$ and $\sigma_{\rm{D}}$[Fe/H] $=0.25$.

Since we used exactly the same spectra as \cite{Sou1998} the only sources of the differences in the results and the estimated uncertainties are the applied numerical procedures, the selection of lines and the EW measurements. We note that our intrinsic uncertainties in $T_{\rm{eff}}$, $\log g$ are $2.6-3.8$ times smaller than the resulting scatter in the results. We also note that our results for $T_{\rm{eff}}$ are systematically 63~K (3 $\sigma$) larger and our results for $\log g$ are systematically 0.15 dex (2.1 $\sigma$) larger while the [Fe/H] results do not show any systematical effects as compared with \citet{Sou1998}. The scatter in all results, comparable to uncertainties as determined from \citet{Katz1998}, and especially in [Fe/H], larger than the average uncertainty as estimated by \citet{Katz1998}, suggests that our results are more precise than those of \cite{Sou1998}.
 
\subsection{Tests with \citet{But2006} data}

\begin{figure*}
   \centering
   \includegraphics[angle=-90,width=16cm]{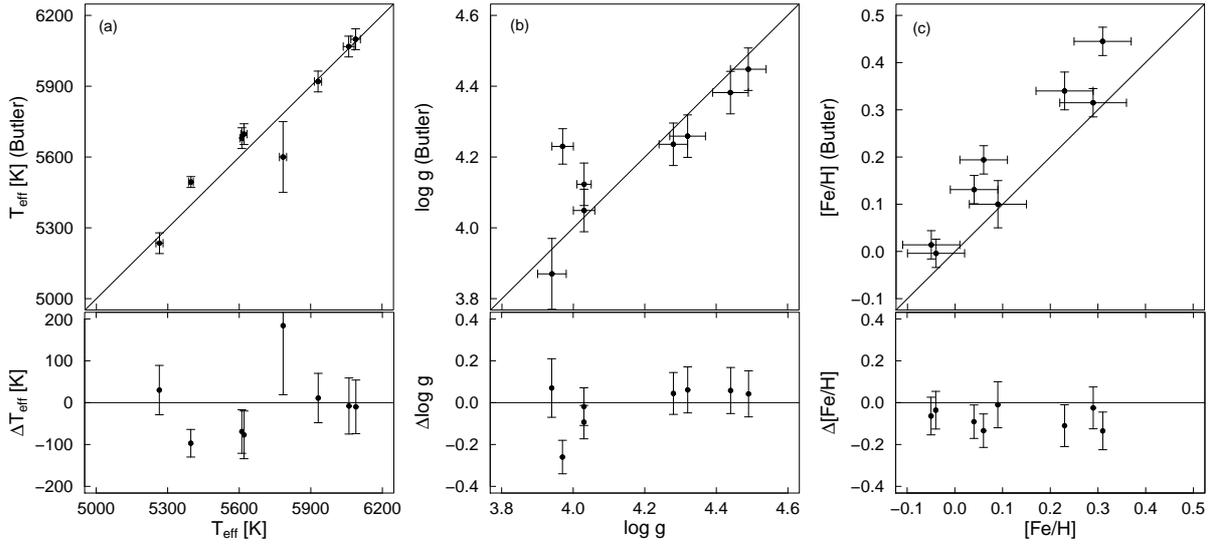}\vspace{-2cm}
      \caption{A comparison of the atmospheric parameters for 8 known planet-hosting stars. Both relations (top) and differences (bottom) between our results and those of \citet{But2006} of $T_{\rm{eff}}$ (a), $\log g$ (b) and [Fe/H] (c) are presented. The one to one relations are presented as solid lines.}
         \label{fig-butler}
   \end{figure*}

We performed another test of quality and reliability of our results using our own HET/HRS spectra of 8 known hosts of planetary systems: HD~10697, HD~38529, HD~72659, HD~74156, HD~75732, HD~88133, HD~118203 and HD~209458, previously analyzed by other authors. The sample contains mainly dwarfs and sub-giants with spectral types ranging from F8 to K0. Our results were compared to the existing determinations collected in \citet{But2006}, who included the nearby stars and exoplanets data from the Lick, the Keck and the Anglo-Australian Observatory planet searches. The results of our analysis for the 8 planet-hosting stars are summarized in Table~\ref{tab-butler}, where also the atmospheric parameters from \citet{But2006} are presented (see also Fig.~\ref{fig-butler}). The mean intrinsic uncertainties for the atmospheric parameters obtained by us and presented in Table~\ref{tab-butler} are $\sigma{T_{\rm{eff}}}=15$~K, $\sigma{\log~g}=0.04$ and $\sigma$[Fe/H] $=0.06$ while the uncertainties presented by \citet{But2006} are: $\sigma{T_{\rm{eff}}}=55$~K, $\sigma{\log~g}=0.06$ and $\sigma$[Fe/H] $=0.03$. The mean intrinsic uncertainties in ${T_{\rm{eff}}}$ as given by the TGVIT are 3.6 times smaller than uncertainties estimated given in \citet{But2006} while our estimated uncertainties in the ${\log~g}$ and the [Fe/H] are comparable. 

After rejecting HD~118203, for which the uncertainty in ${T_{\rm{eff}}}$ as given in \citet{But2006} is 3 times larger than average, we calculated the mean differences and scatter (rms) between our results and those collected in \citet{But2006}. We obtained: $\Delta{T_{\rm{eff}}}=-31$~K, $\Delta{\log~g}=-0.02$, $\Delta$[Fe/H] $=-0.08$ and $\sigma_{\rm{D}}{T_{\rm{eff}}}=49$~K, $\sigma_{\rm{D}}{\log~g}=0.12$ and $\sigma_{\rm{D}}$[Fe/H] $=0.05$.

Our intrinsic uncertainties in $T_{\rm{eff}}$ and $\log g$ are 3.3 and 3 times smaller, respectively, than the resulting scatter in the results. The scatter in [Fe/H] is similar to our uncertainty estimate. 

We note that our results for $T_{\rm{eff}}$ and [Fe/H] are systematically smaller by 31~K (2 $\sigma$) and 0.08~dex (1.3 $\sigma$), respectively, while the results for $\log g$ do not show any measurable systematical effects as compared with \citet{But2006}. The scatter among all results, comparable to uncertainties as determined from \citet{But2006} and ours, and especially in [Fe/H], where it is smaller than our estimate, suggests that the results presented in \citet{But2006} are of similar quality or more accurate than ours.

\begin{table*}
\caption{Atmospheric parameters taken from \citet{But2006} and our results for 8 planet-hosting stars.}
\label{tab-butler} 
\centering 
\begin{tabular}{ll|rrr|rrr}
\hline\hline 
& & \multicolumn{3}{c|}{Butler et al.} & \multicolumn{3}{c}{our results} \\
Name & Sp. & $T_{\rm{eff}}$ & $\log g$ &  [Fe/H] & $T_{\rm{eff}}$ & $\log g$ & [Fe/H] \\
& Type & [K] & [cm s$^{-2}$] &  & [K] & [cm s$^{-2}$] &  \\
\hline 
HD 10697	& G5IV & 5680 $\pm$ 44 &   4.12 $\pm$ 0.06 &  0.19 $\pm$ 0.03 &	5611 $\pm$ 8 &  4.03 $\pm$ 0.02 &  0.06 $\pm$ 0.05 \\
HD 38529	& G4V  & 5697 $\pm$ 44 &   4.05 $\pm$ 0.06 &  0.45 $\pm$ 0.03 &	5620 $\pm$ 13 &  4.03 $\pm$ 0.03 &  0.31 $\pm$ 0.06 \\
HD 72659	& G0   & 5920 $\pm$ 44 &   4.24 $\pm$ 0.06 & $-$0.01 $\pm$ 0.03 &	5931 $\pm$ 15 &  4.28 $\pm$ 0.04 & $-$0.04 $\pm$ 0.06 \\
HD 74156	& G0   & 6068 $\pm$ 44 &   4.26 $\pm$ 0.06 &  0.13 $\pm$ 0.03 &	6060 $\pm$ 23 &  4.32 $\pm$ 0.05 &  0.04 $\pm$ 0.05 \\
HD 75732	& G8V  & 5235 $\pm$ 44 &   4.45 $\pm$ 0.06 &  0.32 $\pm$ 0.03 &	5265 $\pm$ 15 &  4.49 $\pm$ 0.05 &  0.29 $\pm$ 0.07 \\
HD 88133	& G5   & 5494 $\pm$ 23 &   4.23 $\pm$ 0.05 &  0.34 $\pm$ 0.03 &	5397 $\pm$ 10 &  3.97 $\pm$ 0.03 &  0.23 $\pm$ 0.06 \\
HD 118203	& K0   & 5600 $\pm$ 150 &  3.87 $\pm$ 0.10 &  0.10 $\pm$ 0.03 &	5784 $\pm$ 15 &  3.94 $\pm$ 0.04 &  0.09 $\pm$ 0.06 \\
HD 209458	& F8   & 6099 $\pm$ 44 &   4.38 $\pm$ 0.06 &  0.01 $\pm$ 0.03 &	6089 $\pm$ 20 &  4.44 $\pm$ 0.05 & $-$0.05 $\pm$ 0.06 \\
\hline 
\end{tabular}
\end{table*}

To conclude our tests we note that in general the $T_{\rm{eff}}$, $\log g$ and [Fe/H] obtained with the TGVIT and those selected from \citet{Sou1998} or \citet{But2006} agree with each other. The uncertainties in the metallicities, estimated in the TGVIT from the actual Fe abundance distribution as the standard deviation of the mean, are realistic. Our [Fe/H] determinations agree with those of \citet{Sou1998} but are systematically 1.3 $\sigma$ lower than those presented in \citet{But2006}.

The intrinsic TGVIT uncertainties of $T_{\rm{eff}}$ are most certainly underestimated. The comparison with \citet{Sou1998} results suggests that we should enlarge our $\sigma{T_{\rm{eff}}}$ estimates by a factor of 2.6. Similar analysis of scatter between our and \citet{But2006} data suggests that our uncertainties are underestimated by a factor of 3.3. If to apply such corrected $\sigma_{\rm{C}}{T_{\rm{eff}}} = 3.3~\sigma{T_{\rm{eff}}}$ estimate our results agree with those of \citet{Sou1998} or \citet{But2006} within 1~$\sigma_{\rm{C}}$. Similar analysis of scatter leads to a conclusion that our $\sigma{\log~g}$ estimates should be increased by a factor of 3. With such uncertainty estimate our ${\log~g}$ determinations agree with those of \citet{Sou1998} or \citet{But2006} well within 1~$\sigma_{\rm{C}}{\log~g} = 3.0~\sigma{\log~g}$ as well. Since the microturbulence velocity uncertainty is estimated in the same manner as for $T_{\rm{eff}}$ and $\log g$ we expect that these uncertainties are also underestimated by a factor of 3.

\section{Atmospheric parameters}

The results of our determinations of the atmospheric parameters for 348 GK giant stars are presented in Table~\ref{tab-res} (columns 2-5). In all cases the values of $T_{\rm{eff}}$, $\log g$, $v_{\rm{t}}$ and [Fe/H] are presented together with their intrinsic uncertainties. For 332 objects we obtained consistent solutions for each of stellar parameters in (typically) 10 iterations of the procedure described in Sect.~4. 

The spectroscopic analysis revealed that 5 of our presumably giants are actually dwarfs (TYC 0435-01209-1, TYC 0683-01063-1, TYC 1496-01002-1, TYC 3012-00273-1 and TYC 4444-00200-1) with $\log~g \ge 4.0$. On the other hand, 4 objects (TYC 2818-00449-1, TYC 2818-00504-1, TYC 2823-01028-1 and TYC 2823-01398-1) recognized in the SIMBAD as dwarfs are actually K-type giants. 

\addtocounter{table}{5}

The values of $T_{\rm{eff}}$, $\log g$, $v_{\rm{t}}$ and [Fe/H] for 332 star were found to stay generally within the range of the TGVIT model grids. Analysis of 24 stars resulted with effective temperatures below 4250~K with other parameters within the TGVIT range of applicability, however. As the \citet{Kur1993a} models included in the TGVIT reach down to 4000~K we consider these determinations as realistic but associated with larger uncertainties as it will be illustrated in Sect.~6.1. For 16 stars the resulting atmospheric parameters were either incoherent or outside the range of applicability of the TGVIT. For these stars a simplified analysis was performed, as described in Sect.~6.2, and they will be discussed separately. 

\begin{figure*}
   \centering
   \includegraphics[angle=0,width=16.5cm]{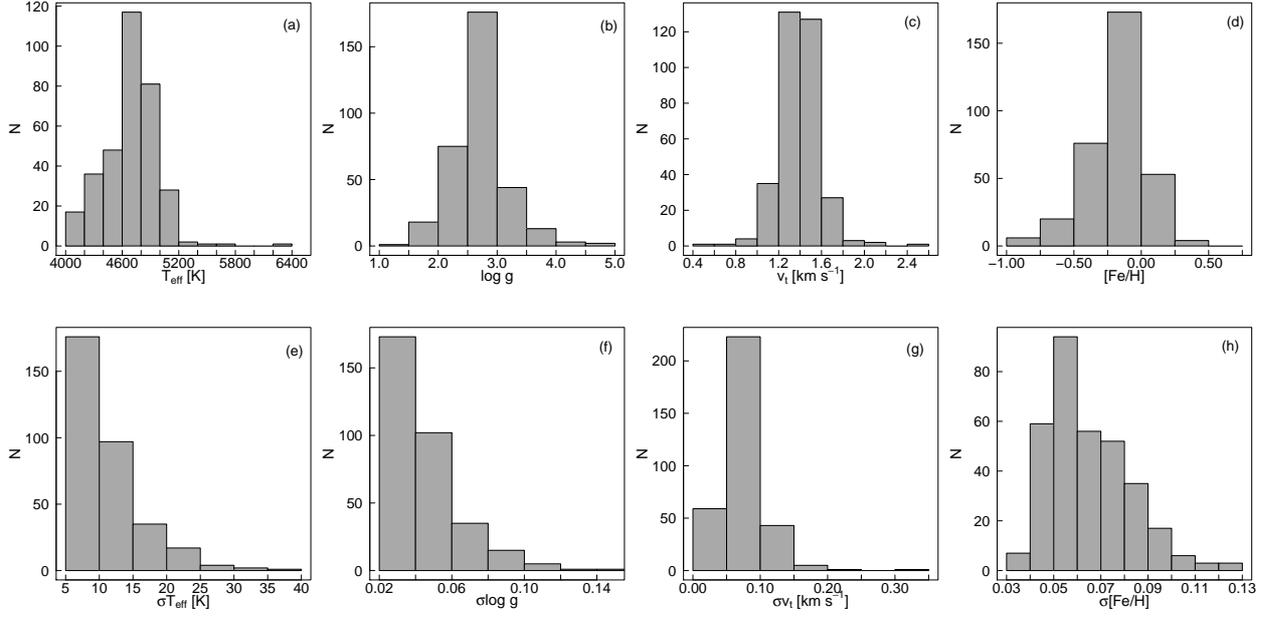}
      \caption{Distributions of atmospheric parameters for the 332 stars with the complete spectroscopic analysis: $T_{\rm{eff}}$, $\log g$, $v_{\rm{t}}$ and 
[Fe/H] (panels a-d) and their intrinsic uncertainties (panels e-h) are presented.}
         \label{fig-atm}
   \end{figure*}

The obtained $T_{\rm{eff}}$ ranges between 4055~K and 6239~K with a median value at 4736~K. The distribution of $T_{\rm{eff}}$ is presented in Fig.~\ref{fig-atm}a. Virtually all our stars have $T_{\rm{eff}}$ between 4000~K and 5200~K (with vast majority, $\sim$~200 of them, between 4600~K and 5000~K) making them mostly G8-K2 stars. The 5 dwarfs are among the hottest stars in our sample with $T_{\rm{eff}}$ between 4906~K and 6239~K. The intrinsic uncertainty distribution in $T_{\rm{eff}}$ is presented in Fig.~\ref{fig-atm}e.

The derived $\log g$ for our sample stars ranges between 1.39 and 4.78 with the median of 2.66. The distribution of $\log g$ is presented in Fig.~\ref{fig-atm}b. 
The majority of our stars (251 of them) have $\log g$ of $2.0-3.0$ making them generally giants. Small fractions of 19 bright giants with $\log~g \le 2.0$ and 3 subgiants with $3.7 \le \log~g \le 4.0$ are present in our sample as well. The intrinsic uncertainty distribution in $\log g$ is presented in Fig.~\ref{fig-atm}f.

The resulting microturbulence velocity, $v_{\rm{t}}$, reaches values from 0.57~km~s$^{-1}$ to 2.49~km~s$^{-1}$ and has the median value of 1.4~km~s$^{-1}$. Most of our stars have $v_{\rm{t}}$ between 1.2~km~s$^{-1}$ and 1.6~km~s$^{-1}$ (Fig.~\ref{fig-atm}c). The value of $v_{\rm{t}}$ for 5 dwarfs in our sample is scattered over the range $0.57-1.68$~km~s$^{-1}$. The intrinsic uncertainty distribution of $v_{\rm{t}}$ is presented in Fig.~\ref{fig-atm}g.

The metallicity of stars in our sample, [Fe/H], stays within $-1.0$ to $+0.45$ range with the median value of $-0.15$. Fig.~\ref{fig-atm}d presents the distribution and shows that our stars are generally less metal abundant than the Sun, most of them have the [Fe/H] in the range of $-0.5$ to $0.0$. The intrinsic uncertainty distribution in [Fe/H] is presented in Fig.~\ref{fig-atm}h.

\begin{figure*}
   \centering
   \includegraphics[angle=-90,width=15.5cm]{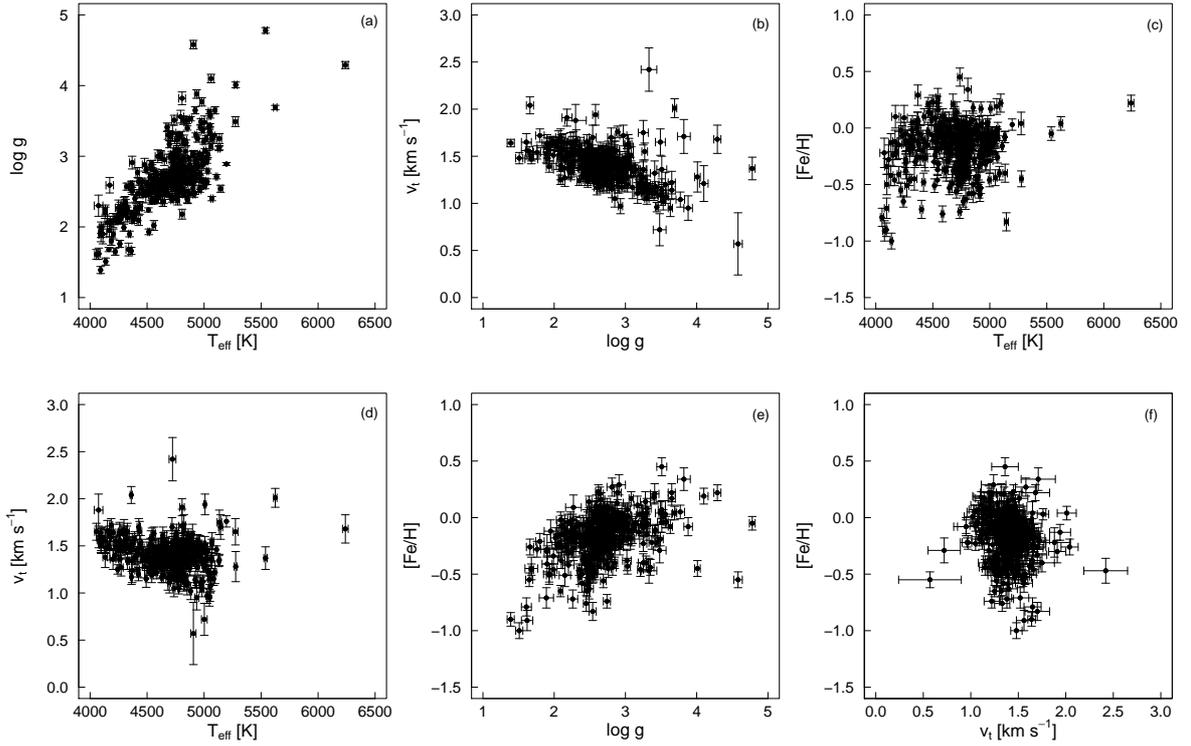}
      \caption{Relations between $T_{\rm{eff}}$, $\log g$, $v_{\rm{t}}$ and [Fe/H] for the 332 stars with the complete spectroscopic analysis.}
         \label{fig-tgmv}
   \end{figure*}

Fig.~\ref{fig-tgmv} presents a few relations between purely spectroscopic atmospheric parameters presented in Table~\ref{tab-res}. They illustrate a general resemblance of our sample to those studied by others (cf. for example \citealt{Tak2008}). In Fig.~\ref{fig-tgmv}a one can notice a clear dependance of $\log g$ on $T_{\rm{eff}}$. The $\log g$ tends to be higher for hotter stars. For stars with $T_{\rm{eff}} \gtrsim 5000$~K the dispersion in surface gravity increases with effective temperature as well. Noticeable relations exist between $v_{\rm{t}}$ and both $\log g$ and $T_{\rm{eff}}$ (Fig.~\ref{fig-tgmv}b and d). The $v_{\rm{t}}$ decreses with both $\log g$ and $T_{\rm{eff}}$. In general [Fe/H] values seem to be significantly lower for low $\log g$ stars (Fig.~\ref{fig-tgmv}e) in our sample, while they reveal uniform distributions in $T_{\rm{eff}}$ (Fig.~\ref{fig-tgmv}c) and $v_{\rm{t}}$ (Fig.~\ref{fig-tgmv}f).


\subsection{Uncertainty estimates}

\begin{figure}
   \centering
   \includegraphics[angle=-90,width=9.5cm]{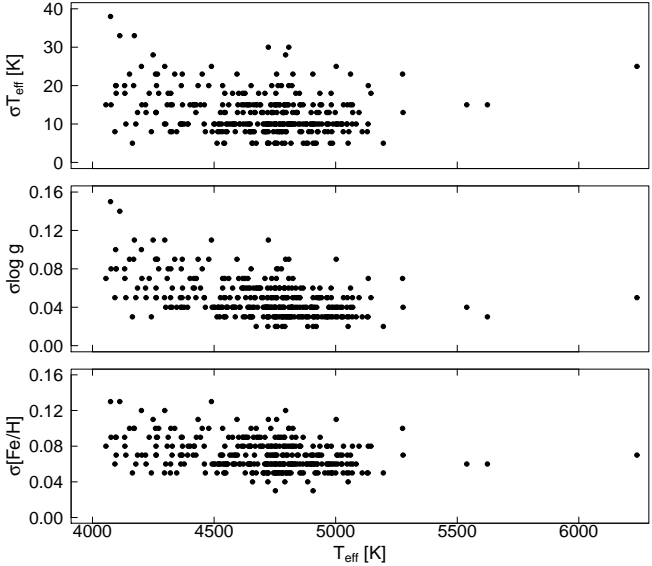}
      \caption{Relations between the intrinsic uncertainties of $\sigma{T_{\rm{eff}}}$ (top panel), $\sigma{\log~g}$ (middle panel) and $\sigma$[Fe/H] (bottom panel) vs. $T_{\rm{eff}}$ for 332 PTPS stars.}
         \label{fig-sigpar}
   \end{figure}

The mean intrinsic uncertainties of our determinations were found to be: $\sigma{T_{\rm{eff}}}=13$~K, $\sigma{\log~g}=0.05$, $\sigma{v_{\rm{t}}}=0.08$~km~s$^{-1}$ and $\sigma$[Fe/H]~$=0.07$. As already discussed in Sect.~5 all these uncertainties, except [Fe/H], are numerical uncertainties resulting from the iterative procedures of the TGVIT and the real uncertainties are factor of 3 larger (since in a comparison with a larger reference sample our intrinsic uncertainty scaling may change we present our intrinsic uncertainties instead of the scalled ones). In the case of [Fe/H] the numerical uncertainties are the standard deviation of the mean Fe\,I and Fe\,II abundances, hence realistic. The intrinsic uncertainties distributions are preseneted in Fig.~\ref{fig-atm} (panels e-h). There exists no correlation between the intrinsic uncertainties and the obtained parameter values for any of the atmospheric parameters. The scatter of the intrinsic uncertainties is also uniformly distributed over the respective parameters values but it is worth to note that for the stars with $T_{\rm{eff}}$ below 4250~K the intrinsic uncertainties are higher, especially in $\log g$ and [Fe/H]. In Fig.~\ref{fig-sigpar} the intrinsic uncertainties of 
$T_{\rm{eff}}$, $\log g$ and [Fe/H] as a function of $T_{\rm{eff}}$ are presented. 

Several stars from the sample demonstrate uncertainties of atmospheric parameters at least two times larger than the mean values. In the case of [Fe/H] we noted intrinsic uncertainties larger than 0.12 for six stars (TYC 0870-00241-1, TYC 1425-01506-1, TYC 3012-01520-1, TYC 3304-00408-1, TYC 4006-00980-1 and TYC 4428-01582-1). These result from the highest scatter of Fe abundances with respect to the excitation potential and EWs in the whole sample.

Five objects (TYC 0096-00005-1, TYC 1425-01506-1, TYC 3011-00791-1, TYC 3304-00408-1 and TYC 3930-01790-1) have significantly larger $T_{\rm{eff}}$ intrinsic uncertainties ($\ge 30$~K). In the case of $\log g$ larger than typical intrinsic uncertainty ($> 0.1$) was obtained for nine stars (TYC 0096-00005-1, TYC 0276-00507-1, TYC 0870-00241-1, TYC 1425-01506-1, TYC 3012-01520-1, TYC 3304-00405-1, TYC 3304-00408-1, TYC 3930-01790-1 and TYC 4006-00980-1). The precision in $v_{\rm{t}}$ is the worst (i.e. $\ge 0.15$~km~s$^{-1}$) for eight objects (TYC 0683-01063-1, TYC 0870-00084-1, TYC 1425-01506-1, TYC 1496-01002-1, TYC 3011-00791-1, TYC 3012-00273-1, TYC 3930-01790-1 and TYC 4444-00200-1).

The reason for such uncertainty increase is not always clear. For some stars it may be associated with low effective temperature, close to the TGVIT model grids limit of $T_{\rm{eff}} = 4250$~K (TYC 0096-0005-1, TYC 1425-01506-1, TYC 3304-00405-1, TYC 3304-00408-1, TYC 0276-00507-1, TYC 3012-01520-1). In several other cases relatively the low signal to noise ratio of our spectra might be the reason (TYC 3011-00791-1, TYC 4006-00980-1, TYC 0870-00084-1, TYC 1496-01002-1, TYC 3012-00273-1, TYC 4444-00200-1, TYC 4428-01582-1).

It is also possible that the larger than expected uncertainties may be due to an unresolved stellar companion. This is the case of TYC 0870-00241-1, a star with excellent quality spectra, and all atmospheric parameters well within the TGVIT range, that after several epochs of the PTPS RV monitoring appeared to have a stellar-mass companion (Niedzielski et al. - in prep.).

\subsection{Stars with incomplete data}

Sixteen stars of our sample have the atmospheric parameters difficult to derive using \citet{Tak2005a} method. For these stars the solutions found with the TGVIT were either inconsistent or unrealistic. To include at least partly such stars in our analysis we adopted $T_{\rm{eff}}$, $\log g$ and initial luminosity estimates from Adam\'ow et al. (in prep.) for all of them. Effective temperatures were calculated from the empirical calibration of \citet{RM2005} based on the Tycho and the 2MASS photometry. Rough estimates of $\log g$ were obtained using the results of \citet{Bil2006} and \citet{Gel2005}. For all these stars the average value of [Fe/H] obtained for the 332 stars with complete spectroscopic analysis, [Fe/H] $=-0.15$, was assumed (bottom part of Table~\ref{tab-res}). 

This approach revealed that three stars were indeed too cool for the approach of \citet{Tak2005a,Tak2005b}. The remaining 13 did not show any different characteristic in comparison with 332 stars with complete spectroscopic analysis and the reasons why they were not applicable to the TGVIT remain to be clarified. Due to much larger uncertainty of the adopted atmospheric parameters for these stars they will be either discussed separately or ignored in the following sections.

\section{Luminosities, masses, ages and radii}

For 271 stars from our sample no Hipparcos parallaxes are available and for another 4 (TYC 0017-01084-1, TYC 3663-00622-1, TYC 4006-00019-1 and TYC 3304-00090-1) the Hipparcos parallaxes $\pi$ are unreliable ($\sigma_{\pi} \geq \pi$) therefore only estimates of masses and radii based on spectrophotometric parallaxes derived here are actually possible in most cases. For such estimates the bolometric luminosities are required in addition to our $\log g$, $T_{\rm{eff}}$ and metallicities. We calculated the intrinsic color index (B$-$V)$_{\rm{0}}$ and the bolometric corrections $BC_{\rm{V}}$ for our stars from \citet{Alo1999} empirical calibration. Using available photometry and assuming the standard interstellar reddening with $R=3.1$ \citep{Rie1985} we estimated luminosities for 57 stars with usable parallaxes. For 275 objects with unknown or unreliable Hipparcos parallaxes we initially assumed $M_{\rm{V}}$ from \citet{Str1981} empirical calibration and estimated luminosities on that basis. 

Propagation of uncertainty was applied to estimate the luminosity uncertainty. The adopted luminosities, together with their uncertainty estimates are presented in Table~5 (column 11). For the 57 stars with reliable parallaxes the values obtained here were adopted (note that the last column in Table~1 identifies the stars with the Hipparcos parallaxes). For the remaining 275 stars luminosities were constrained better via fitting to evolutionary tracks as described in the following section.

\subsection{Stellar masses and luminosities}

   \begin{figure*}
   \centering
   \includegraphics[angle=0,width=15.5cm]{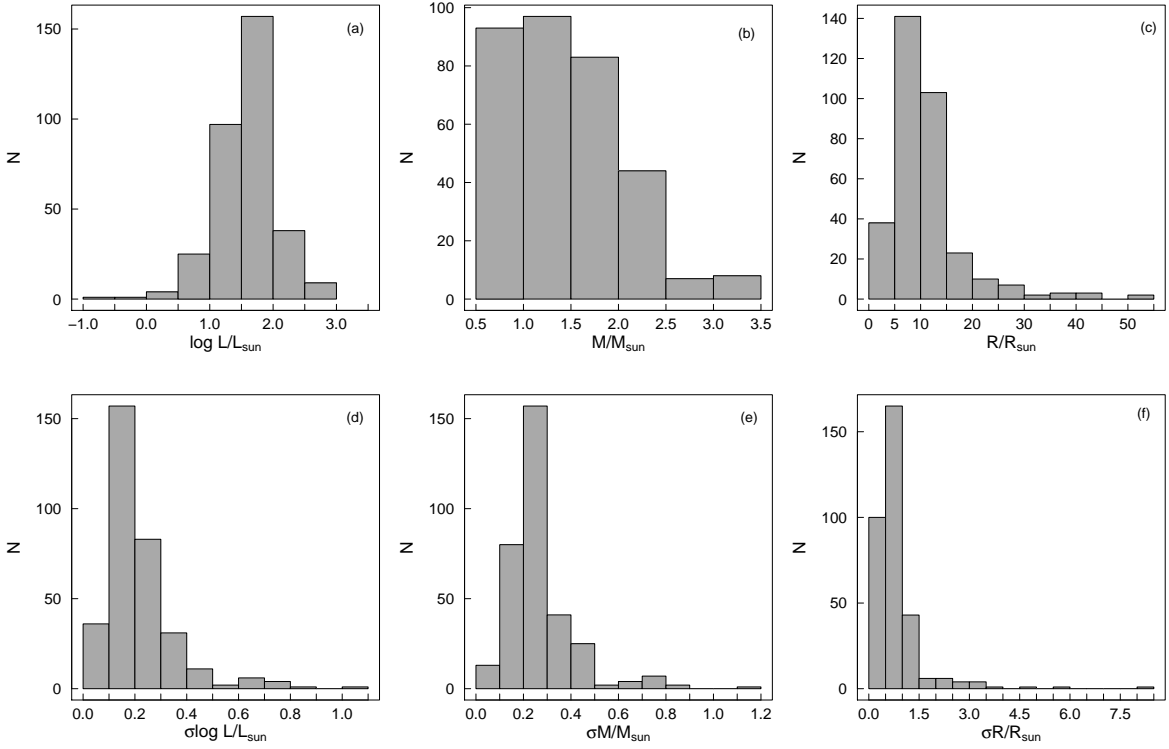}
      \caption{Distributions of luminosities, masses and radii (panels a-c) as well as their uncertainties (panels d-f) for 332 stars with complete spectroscopic analysis.}
         \label{fig-abs}
   \end{figure*}

The stellar masses $M/M_{\rm\odot}$, radii $R/R_{\rm\odot}$ and the estimates of ages were obtained by comparing the positions of stars in the [$\log~L/L_{\rm\odot}$, $\log g$, $\log~T_{\rm{eff}}$] space with the theoretical evolutionary tracks of \citet{Gir2000} and \citet{Sal2000} for a given metallicity (Fig.~\ref{fig-hrd}). We used all available existing tracks corresponding to eight metallicity values: [$Y=0.23, Z=0.0004$], [$Y=0.23, Z=0.001$], [$Y=0.24, Z=0.004$], [$Y=0.25, Z=0.008$], [$Y=0.273, Z=0.019$] (the solar composition), [$Y=0.30, Z=0.03$], [$Y=0.32, Z=0.04$] and [$Y=0.39, Z=0.07$]. To derive $M/M_{\rm\odot}$ for every star the tracks for the nearest metallicity were applied. Then the maximum-likelihood function defined as:

\begin{equation}
\centering
{
\chi^{2} = \sum_{\rm{i=1}}^{n} \left(\frac{ q_{\rm{i}}^{\rm{obs}} - q_{\rm{i}}^{\rm{mod}} }{\sigma_{\rm{i}}}\right)^{2}  ,
}
\label{eq-mlf}
\end{equation}
where $q_{\rm{i}}^{\rm{obs}}$, $q_{\rm{i}}^{\rm{mod}}$ and $\sigma_{\rm{i}}$ denote observed and modeled parameters and their uncertainties respectively, was minimized over the model parameter space. In our case the number of parameters $n$ was set to 3 as $\log~L/L_{\rm\odot}$, $\log g$ and $\log~T_{\rm{eff}}$ were used. The area over which the parameter space was explored was defined as $\pm$ 10 $\sigma$ (intrinsic) in $\log g$ and $\log~T_{\rm{eff}}$ and $\pm$ 3 $\sigma$ in $\log~L/L_{\rm\odot}$. The resulting run of the $\chi^2$ over the model parameters was found to be typically relatively flat, therefore the final stellar mass was calculated as a mean value over an extended area of $\chi^2 \le 3~\chi^{2}_{\rm{min}}$. That procedure resulted in consistent stellar masses determination and realistic estimates of the uncertainties. 

The resulting masses range from 0.6~$M/M_{\rm\odot}$ to 3.4~$M/M_{\rm\odot}$ (see Fig.~\ref{fig-abs}b for a histogram). The majority of stars in our sample have masses below 2~$M_{{\rm\odot}}$. We identified, however, a significant number (63 or $\sim19\%$) of stars which fall in the intermediate-mass range 2~$M_{{\rm\odot}} \le M \le 7~M_{{\rm\odot}}$. Table~\ref{tab-res} (column 12) presents the final adopted masses.

The uncertainty of stellar mass obtained by comparison with the selected here theoretical evolutionary tracks \citep{Gir2000,Sal2000} depends on the accuracy of determination of the position of a star in the [$\log~L/L_{\rm\odot}$, $\log g$, $\log~T_{\rm{eff}}$] space. The effective temperatures and gravities are relatively well determined, the largest source of uncertainty is the parallax, i.e. luminosity. In the case of stars, for which the parallax $\pi$ was precise enough, the metallicity model choice may introduce additional uncertainty since we used the existing evolutionary tracks. 

We found that for an average red giant with known parallax the mass may be estimated within 0.2~$M_{{\rm\odot}}$ in the best cases, while the mean uncertainty for the whole sample is $\sigma{M/M_{\rm\odot}} = 0.3$. The uncertainties may become even larger, $\sim$1~$M_{{\rm\odot}}$, for very confused evolutionary tracks regions at lower effective temperatures ($\log~T_{\rm{eff}} \le 3.65$). 

For most of our stars no reliable parallaxes were available therefore luminosity $\log~L/L_{\rm\odot}$ was constrained better than the initial estimate by adopting the luminosity from the fits to the evolutionary tracks, corresponding to the determined $\log g$, $\log~T_{\rm{eff}}$ and the stellar mass. In Table~\ref{tab-res} (column 11) such luminosity estimate is presented for 275 stars. The $\log~L/L_{\rm\odot}$ ranges from $-0.68$ to $2.86$ with a peak at $1.6$ (Fig.~\ref{fig-abs}a). The mean uncertainty in $\log~L/L_{\rm\odot}$ was found to be $0.19$ in the best cases, but the average value for the whole sample is $\sigma \log~L/L_{\rm\odot} = 0.23$.

The H-R diagram for 332 PTPS stars with the complete spectroscopic analysis is presented in Fig.~\ref{fig-hrd}.

\subsection{Age estimates}

As a large number of presented stars have roughly solar masses, a small uncertainty in mass leeds to a significant uncertainties in the resulting age. On the contrary for the intermediate-mass stars the precision in mass is worse but the estimated ages are more robust. Ages of majority of our program stars are therefore uncertain. The estimated stellar ages are presented in Table~\ref{tab-res} (column 14). For many stars only a wide range of age is given reflecting the complex nature of RGC region and the difficulties in age determination for single field objects. Nevertheless, we found that a typical star from our sample is $3-5$~Gyr old (mean uncertainty in age is $1.1-1.5$~Gyr).

\subsection{Stellar radii}

The stellar radii were estimated from either spectroscopic $\log g$ and the adopted stellar mass or spectroscopic $T_{\rm{eff}}$ and the adopted luminosity. In every case one parameter obtained from the spectroscopic analysis and one from the model fitting was used according to the equations:

\begin{equation}
\centering
{
R/R_{\rm{\odot}} (g, M) = \frac{ \sqrt{G M_{\rm{\odot}} \left(\frac{M}{g}\right)} }{R_{\rm{\odot}}}  ,
}
\label{eq-r1}
\end{equation}
or:

\begin{equation}
\centering
{
R/R_{\rm{\odot}} (T_{\rm{eff}}, L) = \sqrt{ \frac{L}{L_{\rm\odot}}} \left(\frac{ T_{\rm{eff\odot}}}{T_{\rm{eff}} }\right)^{2}  ,
}
\label{eq-r2}
\end{equation}
where $G$ is the gravitational constant and $M_{\rm\odot}$, $R_{\rm{\odot}}$, $L_{\rm{\odot}}$ and $T_{\rm{eff\odot}}$ are the solar values of mass, radius, luminosity and effective temperature.

\begin{figure}
   \centering
   \includegraphics[angle=-90,width=8cm]{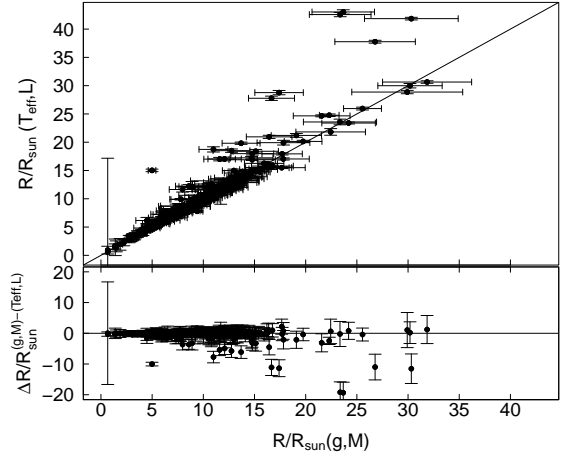}\vspace{-1.5cm}
      \caption{Comparison of radii obtained from two sets of stellar parameters: $\log g$, $M/M_{\rm\odot}$ and $T_{\rm{eff}}$, $\log~L/L_{\rm{\odot}}$ for 332 stars. The solid line corresponds to one to one relation.}
         \label{fig-rrcomp}
   \end{figure}

The maximum uncertainty was estimated in both approaches by application of the logarithmic derivative method. In Fig.~\ref{fig-rrcomp} a comparison of radii derived using both methods is presented. Although for several stars the radii differ significantly, the general agreement between both 
$R/R_{\rm\odot}$ estimates is good. Mean absolute difference between radii determined from the Eqs.~\ref{eq-r1} and \ref{eq-r2} is $\Delta{R}=0.7~R_{\rm{\odot}}$. The largest deviations exist for the most extended stars (cf. Fig.~\ref{fig-rrcomp}). They are most probably a consequence of inaccuracy of the luminosity estimates for those luminous stars without parallax and departure from the LTE conditions in stars with the most extended outer regions. Due to better consistency with the evolutionary status the radii obtained from Eq.~\ref{eq-r2} with their uncertainties are given in Table~\ref{tab-res} (column 13).

The radii of stars in our sample range from 0.6~$R_{\rm{\odot}}$ to 52.1~$R_{\rm{\odot}}$ (Fig.~\ref{fig-abs}c) with majority of about $9-11~R_{\rm{\odot}}$. The mean uncertainty in the derived stellar radii was found to be 0.8~$R/R_{\rm\odot}$ (Fig.~\ref{fig-abs}f). 

\subsection{Stars with incomplete data}

For the sixteen stars with incomplete atmospheric data discussed in Sect.~6.2 and presented in the bottom part of the Table~\ref{tab-res} we applied essentially the same procedure to obtain luminosities, masses, ages and radii. However, as only rough estimates of $\log g$ and assumed [Fe/H] were available the resulting masses, ages and radii are very uncertain. Therefore, we excluded these 16 stars from our discussion presented in Sect.~10.

   \begin{figure}
   \centering
   \includegraphics[angle=-90,width=8cm]{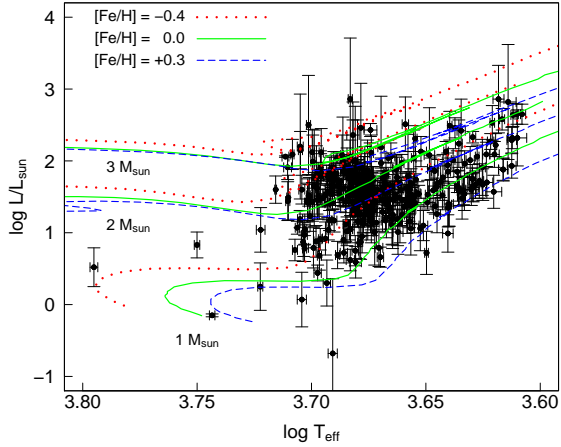}
      \caption{The H-R diagram for 332 PTPS stars with the complete spectroscopic analysis. The theoretical evolutionary tracks \citep{Gir2000} are presented for stellar masses of $1-3~M_{\rm\odot}$ and several metallicities. The dotted, solid and dashed lines correspond to [Fe/H] $=-0.4, 0.0, 0.3$, respectively.}
         \label{fig-hrd}
   \end{figure}

\section{Radial velocities}

\begin{figure}
   \centering
   \includegraphics[angle=0,width=9cm]{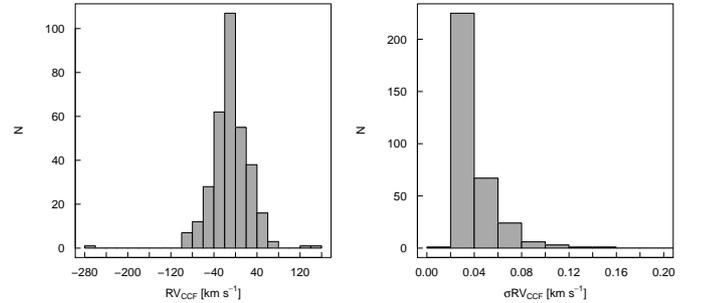}
      \caption{A histograms of the RVs obtained from the cross-correlation function (left panel) and their uncertainties (right panel) for 347 stars.}
         \label{fig-rvcomp}
   \end{figure}

For obtaining absolute radial velocities a cross-correlation analysis with an artificial template was applied. To construct the cross-correlation functions (CCFs) the normalized stellar spectra were correlated with a numerical mask consisting of 1 and 0 value points \citep{Now2008}. The non-zero points corresponded to the positions of 300 non-blended, isolated stellar absorption lines at zero velocity. In our RV measurements only the first 17 orders of the ,,blue'' spectra, free from telluric lines, were used. The CCF was computed step by step for each velocity point without merging the orders. For every order the algorithm selected only lines suitable for a given wavelength range. The CCFs from all orders were finally added to construct the final CCF for the whole spectrum. The final radial velocities were measured by fitting a Gaussian function to the CCF for the whole spectrum. The RVs uncertainties were computed as rms$/\sqrt{17}$ of the 17 RVs obtained from fitting a Gaussian to the CCF for each order seperately. The mean standard uncertainty obtained this way was $\sigma$RV$_{\rm{CCF}} = 0.041$~km~s$^{-1}$. In Fig.~\ref{fig-rvcomp} the distribution of resulting RVs and their uncertainties is presented. The absolute precision of presented RV is lower than the $\sigma$RV$_{\rm{CCF}}$. The HRS is neither thermally nor pressure stabilized therefore the wavelength scale of our spectra is affected by these factors and the real uncertainties are likely to be of the order of 1~km~s$^{-1}$.

In Table~\ref{tab-res} (columns 6-8) the RVs transformed to the barycenter of the Solar System with the algorithm of \citet{Stum1980} for all stars in our sample (except for TYC 3304-00479-1 due to the lack of suitable spectra) are presented together with the epochs of observation ($MJD$). 

\section{Comparison with literature data}

The vast majority of our stars were studied here for the first time and no comparison with literature data is possible. Nevertheless, several individual objects were included in other surveys and we can compare our results with those of other authors, obtained with various methods. 

   \begin{figure}
   \centering
   \includegraphics[angle=-90,width=8cm]{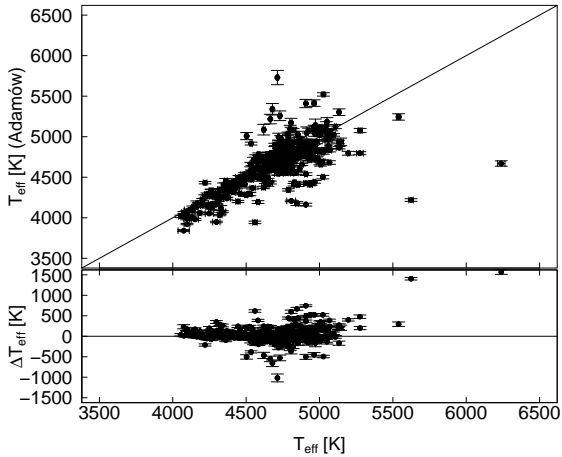}\vspace{-1.5cm}
      \caption{A comparison of $T_{\rm{eff}}$ of the 332 PTPS stars obtained in this study and taken from Adam\'ow et al. (in prep.). Both relation (top) and differences (bottom) are presented. The uncertainties of both determinations are shown. The solid line corresponds to one to one relation.}
         \label{fig-temp}
   \end{figure}
   \begin{figure}
   \centering
   \includegraphics[angle=-90,width=8cm]{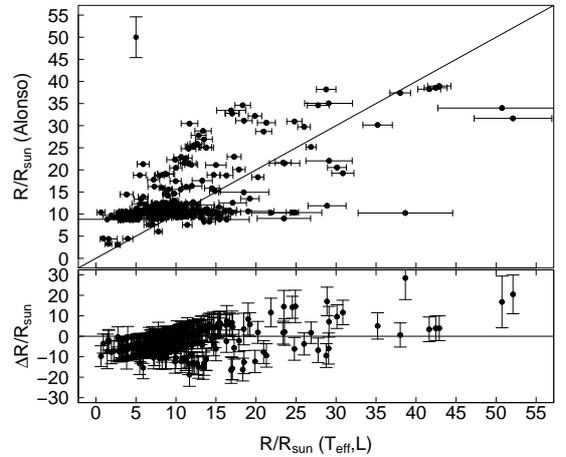}\vspace{-1.5cm}
      \caption{A comparison of the $R/R_{\rm{\odot}}$ for the 332 PTPS stars derived here and calculated from the empirical calibration of \citet{Alo2000}. Both relation (top) and differences (bottom) are presented. The uncertainties in our determinations are shown for every target whereas \citet{Alo2000} a typical uncertainty for 10~$R_{\rm{\odot}}$ star is denoted in the upper left corner. The solid line presents the one to one relation.}
         \label{fig-rrlit}
   \end{figure}
   \begin{figure*}
   \centering
   \includegraphics[angle=-90,width=15.5cm]{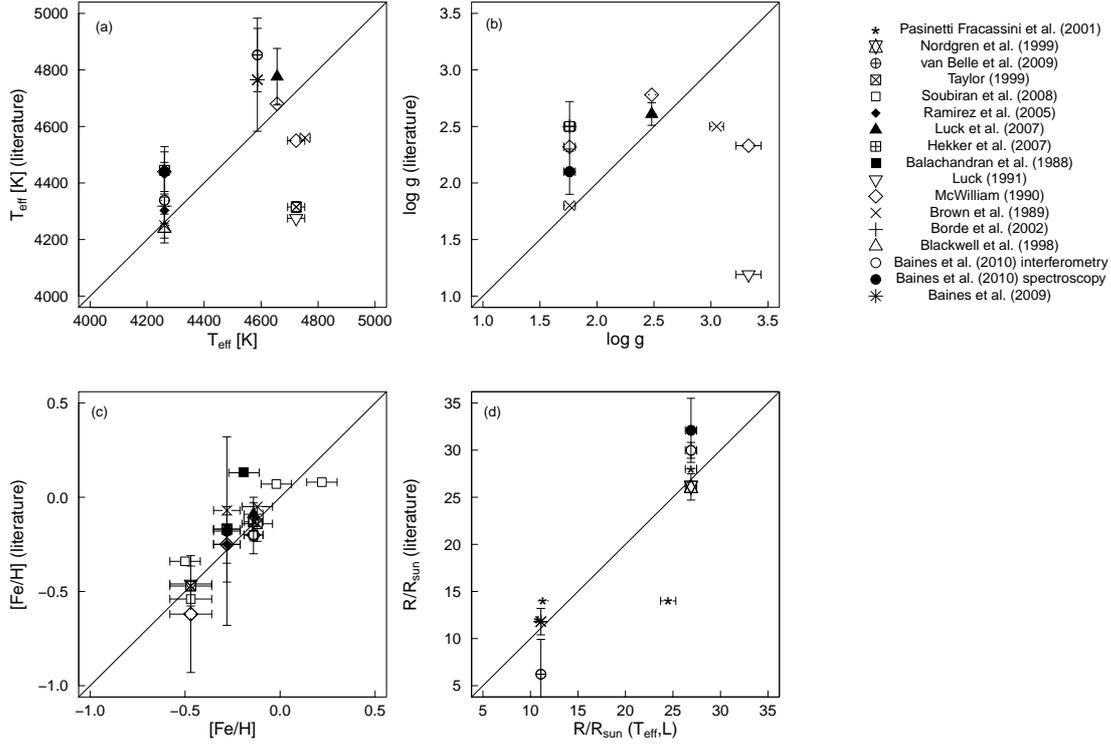}
      \caption{A comparison of various parameters for stars with the literature data. The uncertainties are presented if available. The solid lines denote the one to one relations.}
         \label{fig-parcomp}
   \end{figure*}
   \begin{figure}
   \centering
   \includegraphics[angle=-90,width=8cm]{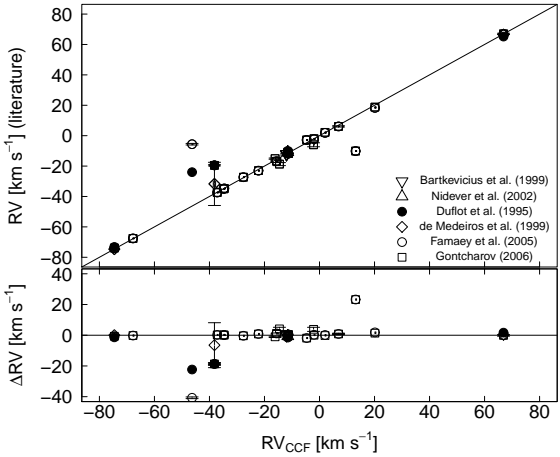}\vspace{-1.5cm}
      \caption{The radial velocities derived from our cross-correlation analysis compared with those available in the literature. Relation (top) and differences (bottom) between the RV results are shown. The RVs uncertainties are presented if available. The solid line presents the one to one relation.}
         \label{fig-rvlit}
   \end{figure}
\begin{table*}
\caption{The mean differences and the standard deviations between stellar parameters presented in this work and derived by other authors.}
\label{tab-comp} 
\centering 
\begin{tabular}{lc|rrrrrc}
\hline\hline 
Reference & Number & $\Delta{T_{\rm{eff}}} \pm \sigma$ & $\Delta{\log~g} \pm \sigma$  &  $\Delta$[Fe/H] $\pm \sigma$ & $\Delta{R} \pm \sigma$ & $\Delta$RV $\pm \sigma$ & Remarks \\
          & of stars & [K] & [cm s$^{-2}$] &  & [$R_{\rm{\odot}}$] & [km s$^{-1}$] & \\
\hline 
Adam\'ow et al. (in prep.)& 332 & 48 $\pm$ 224 & $-$ & $-$ & $-$ & $-$ & ($\dagger$)\\
\citet{Alo2000}			& 332 & $-$ & $-$ & $-$ & $-$2.2 $\pm$ 5.8 & $-$ & ($\dagger$)\\
\citet{Bai2009}			& 1 & $-$178 & $-$ & $-$ & $-$0.7 & $-$ & \\
\citet{Bai2010} spectroscopy	& 1 & $-$179 & $-$0.34 & $-$0.10 & $-$5.2 & $-$ & \\
\citet{Bai2010} interferometry	& 1 & $-$78 & $-$ & $-$ & $-$3.1 & $-$ & \\
\citet{Bal1988}			& 1 & $-$ & $-$ & $-$0.32 & $-$ & $-$ & \\
\citet{Bar1999}			& 1 & $-$ & $-$ & $-$ & $-$ & 0.113 & \\
\citet{Bla1998}			& 1 & 22 & $-$ & $-$ & $-$ & $-$ & \\
\citet{Bor2002}			& 1 & $-$57 & $-$0.74 & $-$ & $-$ & $-$ & \\
\citet{Bro1989}			& 2 & 103 $\pm$ 129 & 0.26 $\pm$ 0.42 & $-$0.14 $\pm$ 0.10 & $-$ & $-$ & ($\dagger$)\\
\citet{Med1999}			& 4 & $-$ & $-$ & $-$ & $-$ & 0.178 $\pm$ 0.169 & ($\dagger$,*)\\
\citet{Duf1995}			& 5 & $-$ & $-$ & $-$ & $-$ & $-$0.405 $\pm$ 1.810 & ($\dagger$,*)\\
\citet{Fam2005}			& 19 & $-$ & $-$ & $-$ & $-$ & 0.163 $\pm$ 0.834 & ($\dagger$,*)\\
\citet{Gon2006}			& 21 & $-$ & $-$ & $-$ & $-$ & 0.422 $\pm$ 1.475 & ($\dagger$,*)\\
\citet{Hek2007}			& 1 & $-$184 & $-$0.74 & $-$0.11 & $-$ & $-$ & \\
\citet{Luck1991}		& 1 & 448 & 2.14 & $-$0.01 & $-$ & $-$ & \\
\citet{LH2007}			& 1 & $-$120 & $-$0.13 & $-$0.05 & $-$ & $-$ & \\
\citet{McW1990}			& 3 & $-$10 $\pm$ 176 & 0.05 $\pm$ 0.84 & 0.06 $\pm$ 0.09 & $-$ & $-$ & ($\dagger$)\\
\citet{Nid2002}			& 1 & $-$ & $-$ & $-$ & $-$ & $-$0.647 & \\
\citet{Nor1999}			& 1 & $-$ & $-$ & $-$ & 0.8 & $-$ & \\
\citet{Pas2001}			& 4 & $-$ & $-$ & $-$ & 1.4 $\pm$ 6.1 & $-$ & ($\dagger$)\\
\citet{RM2005}			& 1 & $-$42 & $-$ & $-$0.03 & $-$ & $-$ & \\ 
\citet{Sou2008}			& 5 & $-$ & $-$ & 0.00 $\pm$ 0.12 & $-$ & $-$ & ($\dagger$)\\
\citet{Tay1999}			& 4 & 408 & $-$ & $-$0.02 $\pm$ 0.06 & $-$ & $-$ & ($\dagger$)\\ 
\citet{Bel2009}			& 1 & $-$266 & $-$ & $-$ & 4.9 & $-$ & \\
\hline 
\end{tabular}
\tablefoot{($\dagger$)~The value after the plus-minus sign is standard deviation (if number of stars to compare is $>$1). (*)~The difference and scatter value obtained with the exception of discrepant stars, see Section 9.1 and Fig.~\ref{fig-rvlit} for details.} 
\end{table*} 

The spectroscopic determinations of $T_{\rm{eff}}$ were compared with effective temperatures derived by Adam\'ow et al. (in prep.) from the empirical calibration of \citet{RM2005} and the Tycho and the 2MASS photometry, where all our stars were included. The comparison of results for 332 stars is shown in Fig.~\ref{fig-temp}. No significant systematic shift is present in the data, but the scatter, of 224~K on average, increases towards higher effective temperatures. The mean difference between our and Adam\'ow et al. (in prep.) temperatures is $\Delta{T_{\rm{eff}}} = 48$~K and only for 4 objects the difference is significantly larger ($ > 740$~K). These stars are: TYC 0683-01063-1, TYC 3319-00170-1, TYC 3930-00665-1 and TYC 3993-01850-1. All of them are significantly reddened by the interstellar extinction what probably affected the photometric $T_{\rm{eff}}$.

Our adopted values of stellar radii for 332 stars were compared with estimates obtained from the empirical calibration of \citet{Alo2000}, based on the angular diameters obtained from the Infra-Red Flux Method and distances computed from the Hipparcos parallaxes. In most cases the results of $R/R_{\rm{\odot}}$ were fairly close to each other and comparable within the estimated uncertainties. It is worth mentioning that uncertainties in radii as estimated from \citet{Alo2000} calibration are large and only a typical uncertainty for a 10~$R_{\rm{\odot}}$ star is presented in Fig.~\ref{fig-rrlit}. The mean difference between our and calibrated radii is $\Delta({R/R_{\rm{\odot}}})=-2.2$ with a scatter of 5.8~$R_{\rm{\odot}}$. It is interesting to note that for the stars with smaller radii our estimates give systematically lower values, while for the stars with larger radii our estimates are systematically larger which is most probably caused by the departure from the LTE.

In Fig.~\ref{fig-parcomp} we present a collective comparison of our parameters for several stars with the data available in the literature. The results obtained by other authors are, in general, in agreement with our determinations of $T_{\rm{eff}}$, $\log g$, [Fe/H] and $R/R_{\rm{\odot}}$. In spite of a very low number of stars in common, i.e. 5 for $T_{\rm{eff}}$, 4 for $\log g$, 8 for [Fe/H] and 5 for $R/R_{\rm{\odot}}$, the agreement is obvious, especially in the case of metallicity. From Fig.~\ref{fig-parcomp} it appears, however, that the scatter between the determinations of various authors is large what makes the comparison more difficult. 

In Fig.~\ref{fig-rvlit} the RVs obtained here using the CCF technique are compared with literature data for all stars, for which such determinations were available. In all but 3 cases our radial velocities agree with those of other authors. Large discrepancies exist in the cases of: TYC 3012-02518-1, TYC 3930-01790-1 and TYC 3304-00090-1. The first two stars are members of binary systems \citep{Abt1981,Mei1968} where the latter is in addition chromospherically active \citep[e.g.][]{Str1994}. The reason for absolute RV discrepancy for TYC 3304-00090-1 is unclear.

The mean differences between the respective parameters are summarized in Table~\ref{tab-comp}. We found that on average our determinations agree with all presented literature results within $\Delta{T_{\rm{eff}}}=-6$~K, $\Delta{\log~g}=0.07$, $\Delta$[Fe/H]$=-0.07$, $\Delta{R/R_{\rm{\odot}}}=-0.6$ and 
$\Delta$RV$=-0.029$~km~s$^{-1}$ (with exception of the 3 discrepant stars). The mean scatter of these comparisons represented by the standard deviation is equal to 179~K, 0.76~dex, 0.10~dex, 3.8~$R_{\rm{\odot}}$ and 0.841~km~s$^{-1}$ for $T_{\rm{eff}}$, $\log g$, [Fe/H] and $R$ and RV, respectively. No systematic effects were found. Our results agree with those of other authors within the estimated uncertainties.

\section{Discussion}

   \begin{figure*}
   \centering
   \includegraphics[angle=-90,width=18cm]{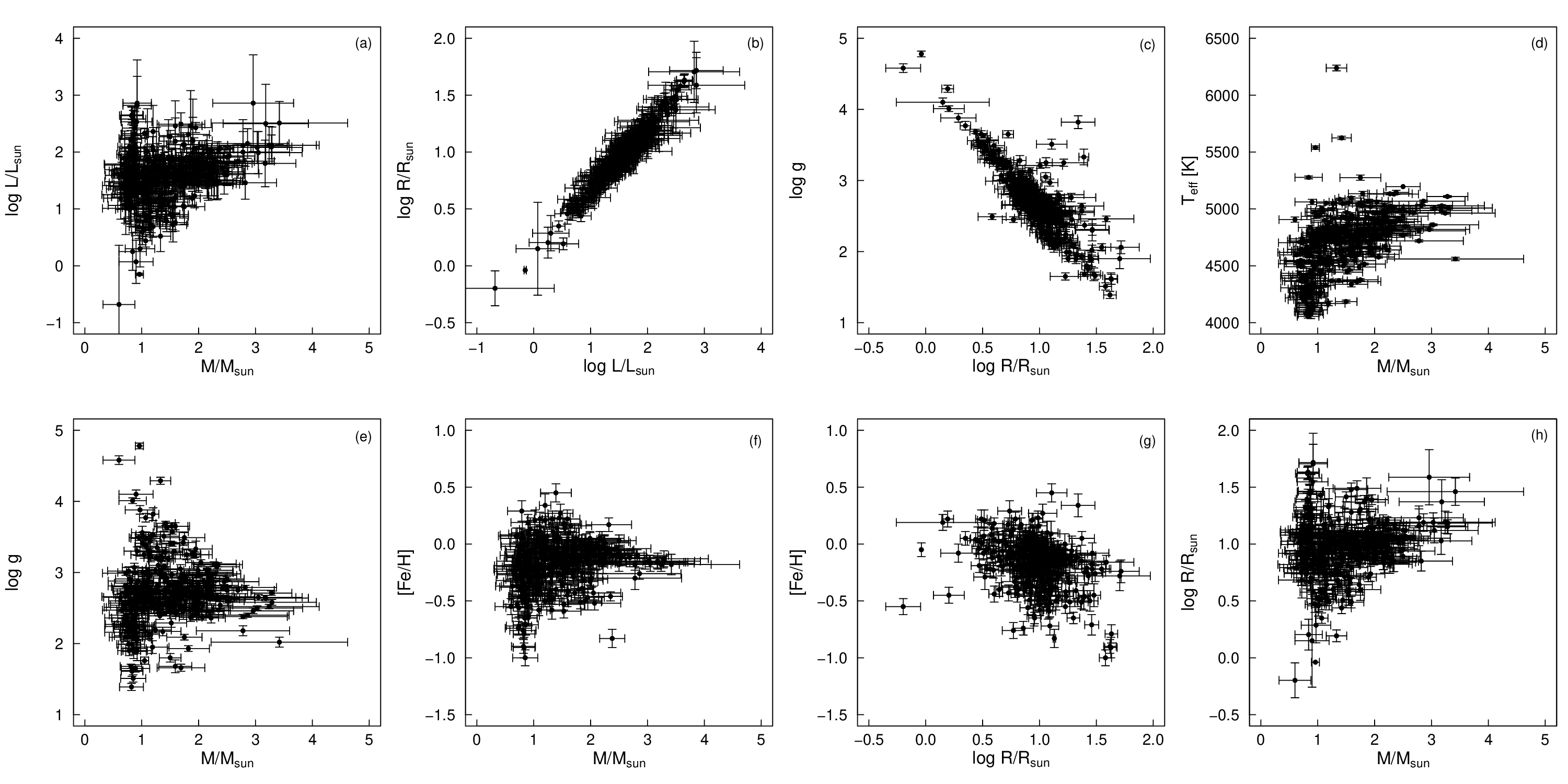}
      \caption{Some relations between parameters for the 332 stars with the complete spectroscopic analysis. In the following panels the relations between $\log~L/L_{\rm{\odot}}$, $M/M_{\rm{\odot}}$, $\log~R/R_{\rm{\odot}}$ and the atmospheric parameters are presented.}
         \label{fig-lmr}
   \end{figure*}
   \begin{figure}
   \centering
   \includegraphics[angle=-90,width=9.5cm]{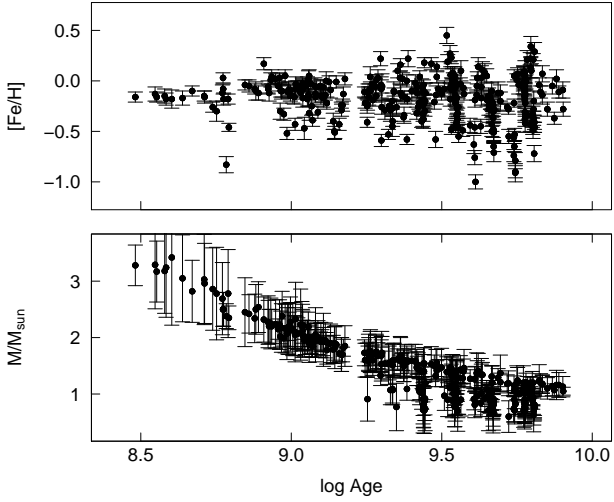}
      \caption{Relations between [Fe/H] and $\log Age$ (top panel) as well as $M/M_{\rm{\odot}}$ and $\log Age$ (bottom panel) for the 332 stars with the complete spectroscopic analysis. The uncertainties in [Fe/H] and $M/M_{\rm{\odot}}$ are only presented. The lower limit of $\log Age$ is presented if in Table~\ref{tab-res} (column 14) the age range is given.}
         \label{fig-amr}
   \end{figure}
   \begin{figure}
   \centering
   \includegraphics[angle=-90,width=7cm]{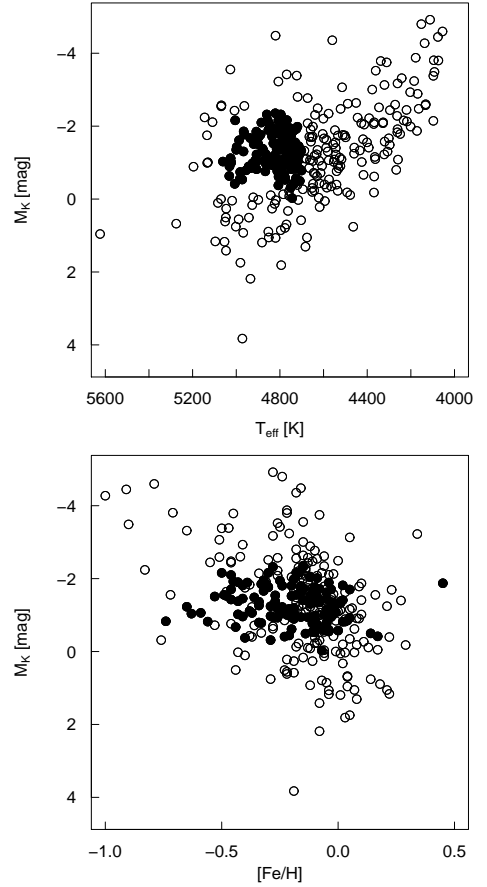}
      \caption{Relations between $M_{\rm{K}}$ and $T_{\rm{eff}}$ (top panel) as well as $M_{\rm{K}}$ and [Fe/H] (bottom panel) for 327 giants. The RGC stars are presented as solid circles while the rest of stars are shown as open circles.}
         \label{fig-Kmagmt}
   \end{figure}

Several relations between the integral and the atmospheric parameters are depicted in Fig.~\ref{fig-lmr}. From the mass-luminosity relation (Fig.~\ref{fig-lmr}a) we infer that relatively high-mass stars tend to be brighter but in the case of solar-mass objects we see a large dispersion in $\log~L/L_{\rm{\odot}}$. A clear correlation exists between $\log~L/L_{\rm{\odot}}$ and $\log~R/R_{\rm{\odot}}$ (Fig.~\ref{fig-lmr}b) reflecting the well-known relation $L \sim R^{2} T_{\rm{eff}}^{4}$. In Fig.~\ref{fig-lmr}c a strong $\log~R/R_{\rm{\odot}}$ dependance on $\log g$, resulting from Eq.~\ref{eq-r1}, is visible. One can also see that the hotter stars in our sample tend to be of higher mass (Fig.~\ref{fig-lmr}d), except for the dwarfs which reveal similar masses for a wide range of effective temperatures ($\sim1300$~K). These effects are to much extend a result of our sample definition well illustrated in $\log~L/L_{\rm{\odot}}$ vs. $\log~T_{\rm{eff}}$ relation (Fig.~\ref{fig-hrd}). In turn, masses of our sample stars show quite uniform distribution versus $\log g$ and [Fe/H] (Fig.~\ref{fig-lmr}e and f). 

In Fig.~\ref{fig-lmr}g we see that our stars are quite uniformly distributed over the metallicity vs. radius plane. Fig.~\ref{fig-lmr}h is similar to the mass-luminosity relation (Fig.~\ref{fig-lmr}a), probably due to the fact that the stellar radii presented in Fig.~\ref{fig-lmr}b result from the derived luminosities. 

The age-metallicity relation for 332 PTPS stars is shown in Fig.~\ref{fig-amr} (top panel). As we stated already in Sect.~7.2 the stellar ages are uncertain and in many cases ambiguously determined. The lower limit of $\log~Age$ is presented if in Table~\ref{tab-res} (column 14) the age range is given. However, the well-known tendency resulting from the Galactic evolution is maintained. In Fig.~\ref{fig-amr} (bottom panel) the relation between stellar age and mass is presented (the Pearson correlation coefficient is $R=-0.937$). For the stars with masses larger than the solar a quadratic fit can be determined: $M/M_{\rm{\odot}} = (1.07\pm0.07)(\log~Age)^2-(21.42\pm1.30)(\log~Age)+(107.93\pm6.04)$. Based in Fig.~\ref{fig-amr} we find that the giants in our sample present a wide range of evolutionary stages. Relatively older stars ($\ge 3.2$~Gyr) have simultaneously lower masses ($\sim1~M_{{\rm\odot}}$) and reveal higher spread in metallicity (0.26). The more massive giants ($>1.5~M_{{\rm\odot}}$), with lower spread in metallicity (0.16), are younger ($< 3.2$~Gyr). 

From Fig.~\ref{fig-hrd} one can see that our sample is composed mainly of regular giants evolving along the RGB and giants from the RGC. Only very few stars appeared to be dwarfs. We applied the definitions of RGC as proposed in \citet{Jim1998} and \citet{Tau2009}, i.e. 4700~K $ \le T_{\rm{eff}} \le $ 5100~K and 1.5 $ \le \log~L/L_{\rm{\odot}} \le $ 1.8 to our data and due to the large uncertainty in our luminosities we extended the $\log~L/L_{\rm{\odot}}$ range by 0.2 to [$1.3-2.0$]. As a result we found 126 stars to fulfill the criteria for the Clump Giants. These stars constitute only about 38$\%$ of our sample. The mean absolute magnitudes of these stars in four bands: $M_{\rm{V}}=(0.985 \pm 0.499)$~mag, $M_{\rm{J}}=(-0.700 \pm 0.502)$~mag, $M_{\rm{H}}=(-1.178 \pm 0.502)$~mag and $M_{\rm{K}}=(-1.282 \pm 0.508)$~mag confirm that they are the Clump Giants. In Fig.~\ref{fig-Kmagmt} we can see our RGC stars together with the rest of the sample in $M_{\rm{K}}$ vs. $T_{\rm{eff}}$ and $M_{\rm{K}}$ vs. [Fe/H] plots. The scatter (defined as the standard deviation) in the absolute magnitudes is large due to the lack of parallaxes and it is not possible to search for relations with either metallicity or temperature. Given the observational uncertainties in luminosity and our sample definition we cannot state that the observed relative frequency of RGC stars reflects the true one.

The stellar masses obtained here by fitting observed stellar parameters to selected evolutionary tracks are model dependent. Since the mass-loss was ignored they also should be considered as upper limits.

\section{Conclusions}

We presented the atmospheric parameters ($T_{\rm{eff}}$, $\log g$, $v_{\rm{t}}$ and [Fe/H]), luminosities, masses, radii, ages and the absolute radial velocities for 348 stars from the RGC sample of the PTPS. For vast majority of them these are the first determinations.

For 332 stars the complete spectroscopic analysis resulted in precise $T_{\rm{eff}}$, $\log g$, $v_{\rm{t}}$ and metallicities. The estimated intrinsic uncertainties in derived parameters are: $\sigma{T_{\rm{eff}}}=13$~K, $\sigma{\log~g}=0.05$, $\sigma{v_{\rm{t}}}=0.08$~km~s$^{-1}$, 
$\sigma$[Fe/H]~$=0.07$ (the uncertainties in effective temperatures, gravities and microturbulence velocities as discussed in Sect.~5 are underestimated by a factor of 3). In case of 16 stars, for which either incomplete data were available or our spectroscopic approach failed, effective temperatures and $\log g$ were estimated from the existing photometric data. 

In turn of our analysis 5 stars from the present sample appeared to be dwarfs and 343 giants, of which 126 are located in the Red Clump. Our RGC sample was found to be composed of stars with $T_{\rm{eff}}$ between 4055~K and 6239~K with the median value at 4736~K (with vast majority, $\sim$200 of them, between 4600~K and 5000~K), generally G8-K2 stars. The determined $\log g$ ranges between 1.39 and 4.78 with the median of 2.66 (majority of our stars, 251 of them, have $\log g$ of $2.0-3.0$) making them generally giants. A small fraction of 19 bright giants with $\log~g \le 2.0$ and 3 subgiants with $3.7 \le \log~g \le 4.0$ is present as well. The microturbulence velocity, $v_{\rm{t}}$, for stars from this sample ranges from 0.57~km~s$^{-1}$ to 2.49~km~s$^{-1}$ and has the median at 1.4~km~s$^{-1}$. The metallicity, [Fe/H], of stars in our RGC sample stays within $-1.0$ to $+0.45$ range with the median of $-0.15$, our stars are generally less metal abundant than the Sun, and most of them have [Fe/H] in the range of $-0.5-0.0$. 

For all 348 stars luminosities, masses, radii and ages were estimated using the presented atmospheric parameters, archive photometric data and the Hipparcos parallaxes when available. The $\log~L/L_{\odot}$ range from $-0.68$ to $2.86$ with the maximum peak at $1.6$. The resulting masses range from 0.6~$M/M_{\odot}$ to 3.4~$M/M_{\odot}$ with the majority of stars having masses below 2~$M_{\rm\odot}$. We identified, however, 63 (or $\sim19\%$) stars which fall in the intermediate-mass range $2~M_{\rm\odot} \le M \le 7~M_{\rm\odot}$. The determined radii range from 0.6~$R_{\rm{\odot}}$ to 52.1~$R_{\rm{\odot}}$. Most of our stars have radii of about $9-11~R_{\rm{\odot}}$. We found that typically stars from our sample are $3-5$~Gyr old and have mean uncertainties in age of around $1.1-1.5$~Gyr. Average uncertainties in derived parameters are $\sigma\log~L/L_{\rm{\odot}}=0.23$, $\sigma{M}=0.3~M_{\rm{\odot}}$, $\sigma{R}=0.8~R_{\rm{\odot}}$.

We find the precision in derived stellar parameters acceptable for the main purpose of our project, i.e. massive planet search. The precision of atmospheric parameters allows for future more sophisticated analysis of individual stars. However, since the lack of the Hipparcos parallaxes was identified as the main source of the uncertainty in luminosities, masses, radii and ages for most of our stars from the presented sample we expect that the GAIA \citep{Lin1994, Per1997, Bai2002} will allow to constrain their intrinsic parameters better.

\begin{acknowledgements}
We thank Dr. Yoichi Takeda as well as Dr. Peter Stetson and Dr. Elena Pancino for making their codes available for us. We thank the HET resident astronomers and telescope operators for their continuous support. We also thank anonymous referee for comments and suggestions that helped us to improve the manuscript. PZ, AN, MA and GN were supported in part by the Polish Ministry of Science and Higher Education grants N N203 510938 and N N203 386237. AW was supported by the NASA grant NNX09AB36G. The Hobby-Eberly Telescope (HET) is a joint project of the University of Texas at Austin, the Pennsylvania State University, Stanford University, Ludwig-Maximilians-Universit\"at M\"unchen, and Georg-August-Universit\"at G\"ottingen. The HET is named in honor of its principal benefactors, William P. Hobby and Robert E. Eberly. The Center for Exoplanets and Habitable Worlds is supported by the Pennsylvania State University, the Eberly College of Science, and the Pennsylvania Space Grant Consortium. This research has made extensive use of the SIMBAD database, operated at CDS (Strasbourg, France) and NASA's Astrophysics Data System Bibliographic Services.
\end{acknowledgements}

%\begin{thebibliography}{}
\bibliographystyle{aa} % style aa.bst
\bibliography{literature} % your references Yourfile.bib
%\end{thebibliography}

% 
\longtab{1}{
%\begin{landscape}
\begin{longtable}{llllcrrrc}
\caption{\label{tab-cat} Basic catalogue data for the sample stars.}\\
\hline\hline
No.	& \multicolumn{3}{c}{Names}    & Spectral	& V	& (B$-$V)  & $\pi$	 & Ref. \\
  &  TYC  &  HD   &   BD   &   Type  &  [mag]  &  [mag]   &   [mas]   &   \\
\hline
\endfirsthead
\caption{continued.}\\
\hline\hline
No.	& \multicolumn{3}{c}{Names}    & Spectral	& V	& (B$-$V)  & $\pi$	 & Ref. \\
  &  TYC  &  HD   &   BD   &   Type  &  [mag]  &  [mag]   &   [mas]   &   \\
\hline
\endhead
\hline
\endfoot
1 &  TYC 0014 00693 1 &  $-$ &  BD+04 126 &  G5 &    9.25 &   0.961	$\pm$	  0.044 &  $-$ &  T	\\
2 &  TYC 0014 00731 1 &  HD 4446 &  BD+04 118 &  G5 &    8.99 &   1.026	$\pm$	  0.035 &  2.00	$\pm$	 1.34 &  H	\\
3 &  TYC 0014 00752 1 &  $-$ &  BD+04 119 &  K0 &    9.75 &   1.146	$\pm$	  0.070 &  $-$ &  T	\\
4 &  TYC 0014 00769 1 &  $-$ &  BD+04 107 &  K0 &   10.11 &   1.326	$\pm$	  0.125 &  $-$ &  T	\\
5 &  TYC 0014 00882 1 &  $-$ &  BD+04 112 &  K2 &    9.96 &   1.307	$\pm$	  0.096 &  $-$ &  T	\\
6 &  TYC 0017 00572 1 &  $-$ &  BD+04 122 &  K2 &    8.97 &   1.132	$\pm$	  0.040 &  2.34	$\pm$	 1.33 &  H	\\
7 &  TYC 0017 00668 1 &  $-$ &  $-$ &  $-$ &   10.46 &   0.749	$\pm$	  0.119 &  $-$ &  T	\\
8 &  TYC 0017 00900 1 &  $-$ &  BD+05 107 &  K0 &    9.67 &   1.036	$\pm$	  0.065 &  $-$ &  T	\\
\end{longtable}
\tablefoot{The following columns present: (1) running number, (2--5) identifications and spectral type from SIMBAD, (6--8) data from Tycho and Hipparcos catalogues (visual magnitude, color index, Hipparcos parallax in microarcseconds), (9) designation of data origin (T, H -- The Hipparcos and Tycho Catalogues \citep{PerESA1997}, K -- \citet{Khar2009}, Tc -- Tycho magnitudes transformed into Johnson magnitudes).}
%\end{landscape}
}

\longtab{5}{
\begin{landscape}
\begin{longtable}{lrrrrrrrrrrrrr}
\caption{\label{tab-res} Derived physical parameters for the PTPS Red Giant Clump sample stars. Intrinsic uncertainties of atmospheric parameters are presented (see Sect.~5 for a discussion).}\\
\hline\hline
No.   &   $T_{\rm{eff}}$   &    $\log g$  &    $v_{\rm{t}}$  &    [Fe/H]  &  RV  &   $MJD$  &  (B$-$V)$_{\rm{0}}$   &  $M_{\rm{V}}$   &   $BC_{\rm{V}}$   &  $\log~L/L_{{\rm\odot}}$  &    $M$ &    $R$ &  Age \\ 
  &  [K]  &  [cm s$^{-2}$]  &  [km s$^{-1}$]  &   & [km s$^{-1}$]  &  +2400000 [d] &    [mag]  & [mag]  &  [mag]   &   &  [$M_{\rm\odot}$]  & [$R_{\rm\odot}$] & [Gyr] \\
\hline
\endfirsthead
\caption{continued.}\\
\hline\hline
No.   &   $T_{\rm{eff}}$   &    $\log g$  &    $v_{\rm{t}}$  &    [Fe/H]  &  RV  &   $MJD$  &  (B$-$V)$_{\rm{0}}$   &  $M_{\rm{V}}$   &   $BC_{\rm{V}}$   &  $\log~L/L_{{\rm\odot}}$  &    $M$ &    $R$ &  Age \\ 
  &  [K]  & [cm s$^{-2}$]   &  [km s$^{-1}$]  &   & [km s$^{-1}$]  &  +2400000 [d] &    [mag]  & [mag]  &  [mag]   &   &  [$M_{\rm\odot}$]  & [$R_{\rm\odot}$] & [Gyr] \\
\hline
\endhead
\hline
\endfoot
  1 &  4820 $\pm$ 10 &  2.47 $\pm$  0.04 &  1.34 $\pm$  0.04 &  $-$0.39 $\pm$   0.06 &     31.133 $\pm$  0.025 &  54792.237534 &  0.946 &  0.82 &  $-$0.328 &  1.91 $\pm$  0.09 &  1.8 $\pm$  0.2 &  13.0 $\pm$   0.4 &   1.2 $\pm$   0.4 \\ 
  2 &  4785 $\pm$ 10 &  2.58 $\pm$  0.03 &  1.38 $\pm$  0.08 &  $-$0.32 $\pm$   0.06 &     31.499 $\pm$  0.024 &  54428.223773 &  0.972 &  0.33 &  $-$0.342 &  1.91 $\pm$  0.66 &  1.5 $\pm$  0.3 &  13.1 $\pm$   2.3 &   2.6 $\pm$   1.7 \\ 
  3 &  4709 $\pm$ 08 &  2.43 $\pm$  0.04 &  1.43 $\pm$  0.04 &  $-$0.34 $\pm$   0.05 &     $-$1.671 $\pm$  0.024 &  54420.247014 &  1.011 &  0.80 &  $-$0.376 &  1.73 $\pm$  0.19 &  1.3 $\pm$  0.3 &  11.0 $\pm$   0.6 &   3.5 $\pm$   1.1 \\ 
  4 &  4540 $\pm$ 05 &  2.35 $\pm$  0.03 &  1.24 $\pm$  0.05 &  $-$0.46 $\pm$   0.05 &     28.285 $\pm$  0.029 &  54786.223617 &  1.096 &  0.76 &  $-$0.462 &  1.60 $\pm$  0.25 &  0.8 $\pm$  0.3 &  10.2 $\pm$   0.8 &   3.6 - 9.3 \\ 
  5 &  4368 $\pm$ 10 &  2.19 $\pm$  0.04 &  1.52 $\pm$  0.07 &  $-$0.04 $\pm$   0.07 &    $-$29.305 $\pm$  0.092 &  54046.260480 &  1.243 & $-$2.50 &  $-$0.584 &  1.85 $\pm$  0.19 &  1.2 $\pm$  0.3 &  14.7 $\pm$   0.8 &   6.7 $\pm$   5.4 \\ 
  6 &  4548 $\pm$ 08 &  2.51 $\pm$  0.03 &  1.28 $\pm$  0.05 &  $-$0.32 $\pm$   0.05 &     21.017 $\pm$  0.027 &  54428.235521 &  1.106 &  0.74 &  $-$0.458 &  1.79 $\pm$  0.58 &  0.7 $\pm$  0.4 &  12.7 $\pm$   2.1 &   3.6 - 9.3 \\ 
  7 &  4757 $\pm$ 15 &  2.69 $\pm$  0.05 &  1.19 $\pm$  0.06 &  $-$0.37 $\pm$   0.06 &    $-$12.755 $\pm$  0.021 &  54812.164960 &  0.980 &  0.80 &  $-$0.354 &  1.44 $\pm$  0.07 &  1.1 $\pm$  0.1 &   7.7 $\pm$   0.2 &   7.5 $\pm$   2.5 \\ 
  8 &  4543 $\pm$ 05 &  2.40 $\pm$  0.03 &  1.28 $\pm$  0.05 &  $-$0.37 $\pm$   0.05 &     20.670 $\pm$  0.025 &  54039.133009 &  1.103 &  0.76 &  $-$0.461 &  1.46 $\pm$  0.30 &  0.7 $\pm$  0.3 &   8.7 $\pm$   0.9 &   4.7 - 9.3 \\
\end{longtable}
\tablefoot{The following columns present: (1) running number, (2--5) atmospheric parameters and their intrinsic uncertainties (effective temperatures, surface gravities, microturbulence velocities and metallicities), (6) radial velocity from CCF, (7) epoch (MJD) of the radial velocity, (8--10) data obtained from \citet{Alo1999} empirical calibration (color index, absolute magnitude and bolometric correction), (11--14) derived stellar integral  parameters and their uncertainties (luminosities, masses, radii and ages).}
\end{landscape}
}

\end{document}